\documentclass[manuscript=article,layout=twocolumn]{achemso}
\UseRawInputEncoding
\usepackage[version=3]{mhchem}
\usepackage{amsmath,amssymb,bm}
\usepackage{graphicx}
\usepackage{xcolor}
\usepackage{physics}
\usepackage{float}
\usepackage{makecell}
\usepackage{inputenc}
\usepackage[normalem]{ulem}

\title{\rev{Substrate-Mediated Evaporation and Stochastic Evolution of Supported Au Nanoparticles}}

\author{Dmitri N. Zakharov}
\email{dzakharov@bnl.gov}
\affiliation{Center for Functional Nanomaterials, Brookhaven National Laboratory, Upton, NY 11973, USA}

\author{Xiaohui Qu}
\affiliation{Center for Functional Nanomaterials, Brookhaven National Laboratory, Upton, NY 11973, USA}

\author{Hong Wang}
\affiliation{Computational Science Initiative, Brookhaven National Laboratory, Upton, NY 11973, USA}

\author{Yuewei Lin}
\affiliation{Computational Science Initiative, Brookhaven National Laboratory, Upton, NY 11973, USA}

\author{Aaron Stein}
\affiliation{Center for Functional Nanomaterials, Brookhaven National Laboratory, Upton, NY 11973, USA}

\author{James P. Horwath}
\affiliation{Department of Materials Science and Engineering, University of Pennsylvania, Philadelphia, PA 19104, USA}

\author{Shinjae Yoo}
\affiliation{Computational Science Initiative, Brookhaven National Laboratory, Upton, NY 11973, USA}

\author{Eric A. Stach}
\affiliation{Department of Materials Science and Engineering, University of Pennsylvania, Philadelphia, PA 19104, USA}

\author{Alexei V. Tkachenko}
\email{oleksiyt@bnl.gov}
\affiliation{Center for Functional Nanomaterials, Brookhaven National Laboratory, Upton, NY 11973, USA}


\newcommand{\rev}[1]{\textcolor{black}{ #1 }}
\newenvironment{rev_par}
  {\begingroup\color{black}}
  {\endgroup}

\begin{document}

\begin{tocentry}
\begin{center}
\includegraphics{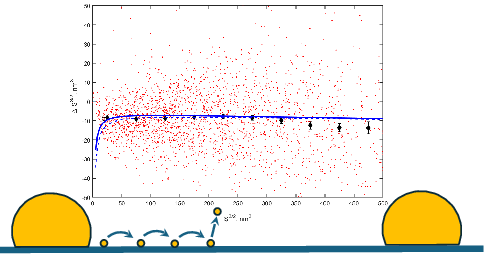}
\end{center}
\end{tocentry}

\begin{abstract}
We use \textit{in situ} transmission electron microscopy with automated tracking to study supported gold nanoparticles (NPs) during high-temperature vacuum annealing. \rev{The average mass loss per NP is governed by a flat, nearly size-independent substrate-mediated evaporation profile.} On top of \rev{this mean shrinkage}, individual NPs show significant fluctuations in apparent growth or shrinkage, and NP volume follows a \rev{random-walk-like trajectory. To rationalize both the ensemble-mean behavior and the particle-resolved variability, we develop a self-consistent theory that couples substrate-mediated evaporation to collective 2D Ostwald-type mass exchange through a shared adatom field, described in terms of a renormalized screening length and background concentration. In the experimentally relevant regime, the theory predicts an approximately size-independent mean shrinkage rate and clarifies how net mass loss suppresses classical coarsening.} \rev{Superimposed on this deterministic drift, we quantify stochastic volume trajectories and capture their fluctuation spectrum with a minimal Langevin description consistent with intermittent adatom attachment and detachment events.} In addition, we characterize the lateral diffusive motion of NPs, which is responsible for their coalescence. Altogether, our results highlight that stochasticity is intrinsic at the nanoscale \rev{and that predicting the evolution of supported NPs at early and intermediate times requires a unified framework combining substrate-mediated evaporation, collective mass exchange, and stochastic fluctuations.}
\end{abstract}

\section{Introduction}

Supported metal nanoparticles (NPs) are cornerstone materials in modern science and technology. Their unique optical, catalytic, and electronic properties make them indispensable in heterogeneous catalysis \cite{Sapi2021_MetalNP_Catalysis,vanDeelen2019_MSI_NatCatal}, plasmonics \cite{NanoscaleHorizons2024_PlasmonicSensors,Sensors2021_LSPR_Review}, and advanced sensing platforms based on localized surface plasmon resonance (LSPR). The remarkable performance of these nanoscale systems is tightly linked to particle size, morphology, and spatial distribution, all of which evolve dynamically under reaction conditions. Control of metal-support interactions is particularly crucial for enhancing activity, selectivity, and stability, and continues to be a major theme in catalyst design \cite{vanDeelen2019_MSI_NatCatal}. However, the same structural features that confer high activity also render NPs unstable, as they undergo thermally and chemically driven processes such as nucleation, migration, coalescence, Ostwald ripening, and sublimation.

\begin{rev_par}
Understanding and predicting this evolution remains a central challenge. While many theoretical descriptions focus on individual mechanisms or asymptotic coarsening behavior, supported nanoparticle systems frequently operate in regimes where multiple pathways—substrate-mediated mass exchange, collective interactions through diffusing adatoms, direct evaporation, and stochastic fluctuations—act simultaneously. Developing a unified framework that consistently connects these microscopic transport processes to ensemble evolution is therefore essential for predictive control of nanoparticle stability.
\end{rev_par}

Our understanding of these mechanisms has been transformed by the rapid development of \textit{in situ} and operando electron microscopy, which enables direct observation of NP dynamics with atomic-scale resolution under realistic environments. Recent reviews highlight how \textit{in situ} TEM has become a central tool for probing sintering, coarsening, and transformation processes in real time \cite{ChemRev2024_OperandoIntro,NanoToday2024_InSituVideo_DL}. Early work with high-resolution TEM established the classical picture of Ostwald ripening, where smaller particles dissolve and larger ones grow through diffusive exchange \cite{SIMONSEN2011}. Subsequent systematic studies categorized nanoparticle sintering into two dominant pathways: Ostwald ripening and particle migration with coalescence (PMC). Hansen et al.\ showed that ripening often dominates initially, while PMC becomes significant once small particles are depleted \cite{Hansen2013}. More recent work has blurred this dichotomy: DeLaRiva et al.\ demonstrated that both pathways can coexist under certain conditions \cite{Delariva2013}, and Visser et al.\ directly imaged Ni NPs undergoing both processes under CO$_2$ hydrogenation, revealing dynamic feedback between discrete coalescence and ripening events \cite{Visser2023}.

Nucleation introduces additional complexity. Although often obscured by temporal and spatial limitations, \textit{in situ} liquid-phase TEM has captured NP nucleation and early stage growth, providing rare experimental evidence that supports classical nucleation theory while underscoring the stochastic nature of cluster formation \cite{Woehl2012}. In contrast to these agglomerative mechanisms, sublimation acts as a competing mass loss pathway, particularly under high-temperature or vacuum conditions. Asoro et al.\ showed that Ag NPs exhibit abrupt, size-dependent sublimation \cite{Asoro2013}, while Li et al.\ revealed that the pathway depends strongly on morphology, proceeding via atomic layer peeling or fragment detachment \cite{Li2019}. Similar considerations apply to Au, where evaporation can induce reshaping and facet evolution \cite{Malyi2012}. More broadly, recent {\it in situ} studies confirm that sublimation and evaporation contribute substantially to nanoparticle degradation \cite{Materials2021_Sublimation_Melting,PRE2013_Au_Evaporation_Pressure}.

The interplay of these mechanisms has been quantified in recent \textit{in situ} studies. Horwath, Voorhees, and Stach demonstrated that even nominally identical supported nanocatalysts can follow divergent trajectories under identical conditions, with coarsening, sintering, and sublimation occurring simultaneously \cite{stach2021}. This variability highlights the inadequacy of mean-field models and the need to incorporate stochasticity explicitly into mechanistic frameworks. In a follow up study \cite{eric2023},  a combination of TEM microscopy and Monte Carlo simulations have been used to elucidated the connection between th NP surface rearrangements and the sublimation kinetics.   Theoretical work by Zinke-Allmang provided an early statistical perspective on nucleation, coalescence, and coarsening \cite{zinke1999}, while more recent models integrate nonequilibrium statistical mechanics and coarse-grained simulations to describe particle reshaping, intermixing, and facet evolution under fluctuating conditions \cite{Lai2019,CES2025_OR_Dimers}. Advances in theory have also revealed novel aspects of Ostwald ripening specific to supported systems, such as the role of dimers and small clusters \cite{CPL2012_NovelOR_Supported}.

A persistent challenge is separating deterministic and stochastic contributions. While traditional models often treat fluctuations as secondary, nanoscale systems are inherently noisy: adatom diffusion, thermal fluctuations, and electron-beam effects all contribute significantly. Beam-induced processes can significantly alter sublimation pathways  \cite{Panciera2019, AuSi2022_Eutectic_Dynamics, Egerton2019_Micron_DamageReview,ActaMat2021_EBeamEffects,MRSBull2020_EBeam_Chemistry}. Thus, a comprehensive description of NP dynamics must explicitly incorporate both deterministic and stochastic components.

In this work, we investigate the stochastic evolution of Au nanoparticles (NPs) supported on Si$_3$N$_4$ under high temperature vacuum annealing.  We employ  high-resolution \textit{in situ} TEM equipped with high temporal resolution direct electron detector. A related systems have been previously studied in refs. \cite{stach2021,eric2023}. Important distinctions  of the current study include: (i) prolonged periods of NP evolution without e-beam exposure, (ii) significant role of NP diffusion and coalescence, and (iii) focus on quantitative modeling of  stochastic effects.
By tracking individual particles, we separate slow, deterministic trends (sublimation and coalescence) from fast, random fluctuations. We find that the net sublimation rate is nearly size independent and several orders of magnitude lower than the Hertz-Knudsen prediction. Under the experimental conditions,   the classical Ostwald ripening is suppressed due to the overall mass loss through substrate mediated evaporation. Superimposed on this flat evaporation profile, nominally similar particles exhibit large variability in apparent growth/shrinkage rate, and their volumes evolve with random walk statistics. We show that a simple model accounts quantitatively and consistently for both the flat evaporation profile and the fluctuation spectrum. It incorporates substrate-mediated evaporation plus a stochastic term representing intermittent adatom attachment/detachment. Finally, we quantify lateral NP diffusion, which sets the collision frequency underlying coalescence. Together, these results show that stochastic processes are not marginal effects but an essential  component of nanoparticle stability and evolution.

\rev{
A complementary and widely used framework for describing particle evolution is provided by population balance models (PBMs), which track the time-dependent size distribution under mechanisms such as evaporation/condensation, Ostwald ripening, and coalescence. Classical and modern PBM formulations incorporate collision kernels, mass-transfer fluxes, and in some cases stochastic event statistics to describe ensemble-level dynamics \cite{Ramkrishna2000_PBM,Friedlander2000_Smoke,Aldous1999_Coagulation,MarchisioFox2013_PBM}. These approaches are particularly powerful in the long-time regime, where systems approach asymptotic scaling laws or self-similar distributions. However, PBMs typically operate at the distribution level and rely on closure assumptions for microscopic fluxes and interaction kernels. In contrast, the present study targets early and intermediate time scales accessible to particle-resolved \textit{in situ} TEM, where the system has not reached an asymptotic coarsening state and where trajectory-to-trajectory variability can be directly measured. Our approach therefore focuses on deriving and validating the microscopic deterministic drift and stochastic noise terms—specifically substrate-mediated evaporation and collective 2D coupling through the adatom field—that underlie ensemble descriptions while explicitly quantifying single-particle fluctuations.
}

\begin{figure}[h!]
\centering   
\includegraphics[width=1\linewidth]{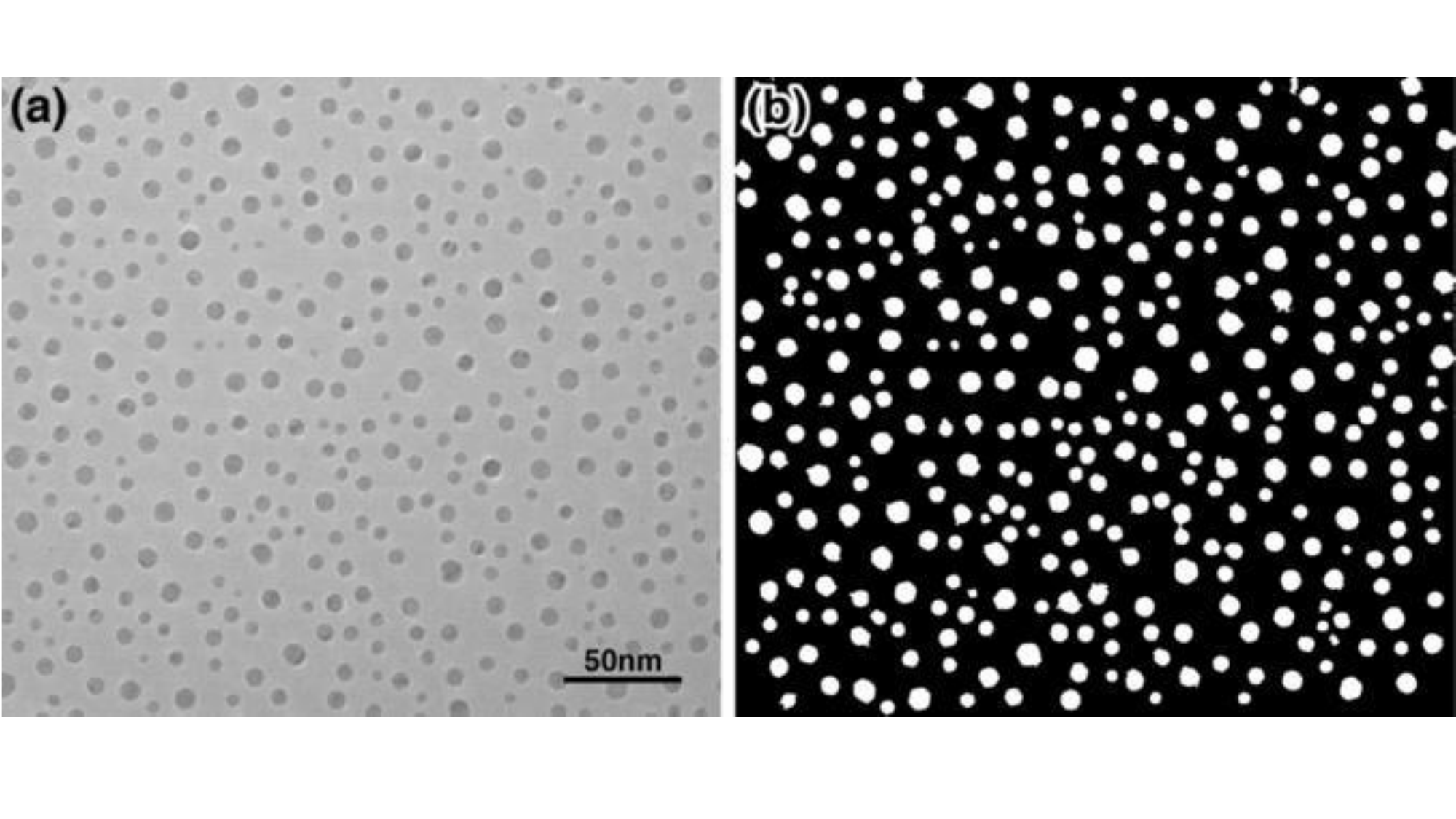}
\caption{ \textbf{(a)} Bright-field TEM image of Au nanoparticles supported on an amorphous Si$_3$N$_4$ membrane during annealing at $950\,^{\circ}\mathrm{C}$ in the microscope column.
  \textbf{(b)} Binary image  generated from (a) by the automatic pipeline used for quantification. Extracted particle footprints (white)  are tracked frame-to-frame for size and position statistics.}
\label{fig:image}
\end{figure}

\section{  Methods }
A film of gold with a nominal thickness of 1.0nm was deposited onto DENS Solutions Wildfire heating Nano-chip with $Si_3N_4$ windows by electron beam evaporation technique in Kurt J. Lesker PVD75 system at $3\times 10^{-6}$ Torr base pressure. As deposited gold film was then annealed at $950^oC$ for over 120 minutes in a column of Environmental Cs-corrected FEI Titan 80-300 (S) TEM operated at 300kV with an information limit of 0.1nm.  The base pressure in the microscope column near the sample was $2\times 10^{-7}$ Torr. The microscope is equipped with Gatan direct electron detector capable of acquiring streams of 1920 x 1792 pixels images at 400 frames per second. The automatic particle indexing procedure \cite{gonzales2006} was applied to the sequence of TEM images (Figure \ref{fig:image}a) to extract particle sizes and positions for progressive time frames (Figure \ref{fig:image}b). Thus, each NP is represented by a time-dependent image of its footprint.  

 \begin{figure}[h!]
\centering   
\includegraphics[width=1.1\linewidth]{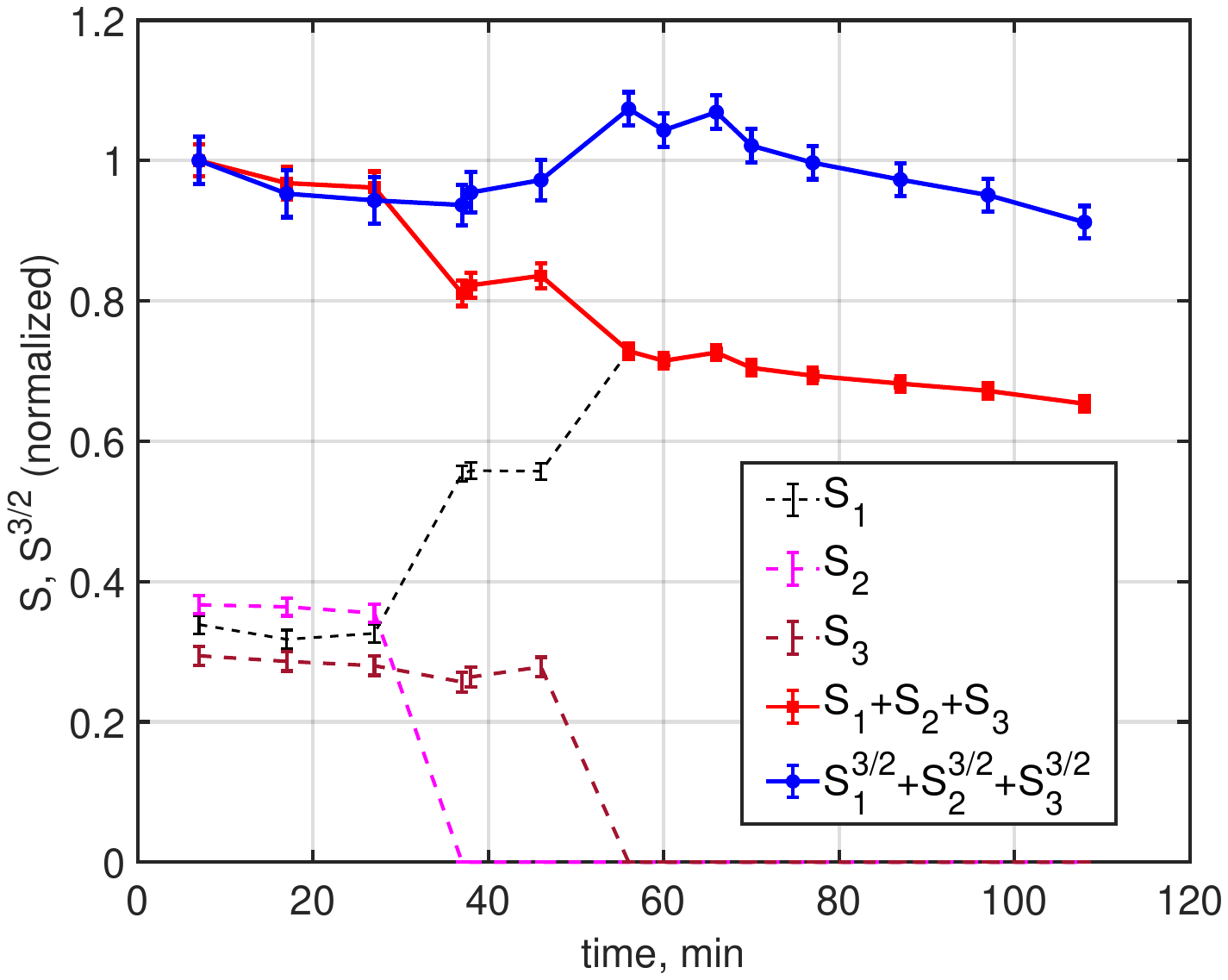}
\caption{\textbf{Volume-conserving coalescence.}
  Evolution of three neighboring Au NPs undergoing two consecutive merger events.  The total footprint area,  $S_1+S_2+S_3$, is not conserved. In contrast, the  total volume proxy parameter,  $S_1^{3/2}+S_2^{3/2}+S_3^{3/2}$, stays approximately constant.\rev{ $S$ and $S^{3/2}$ are normalized so that both sum to $1$ at $t=7$ min}. 
}\label{fig:merge}
\end{figure}

\section{Results and Discussion}
\subsection{Dynamics of NP number and size}

\begin{figure}[h!]
\centering   
\includegraphics[width=1.1\linewidth]{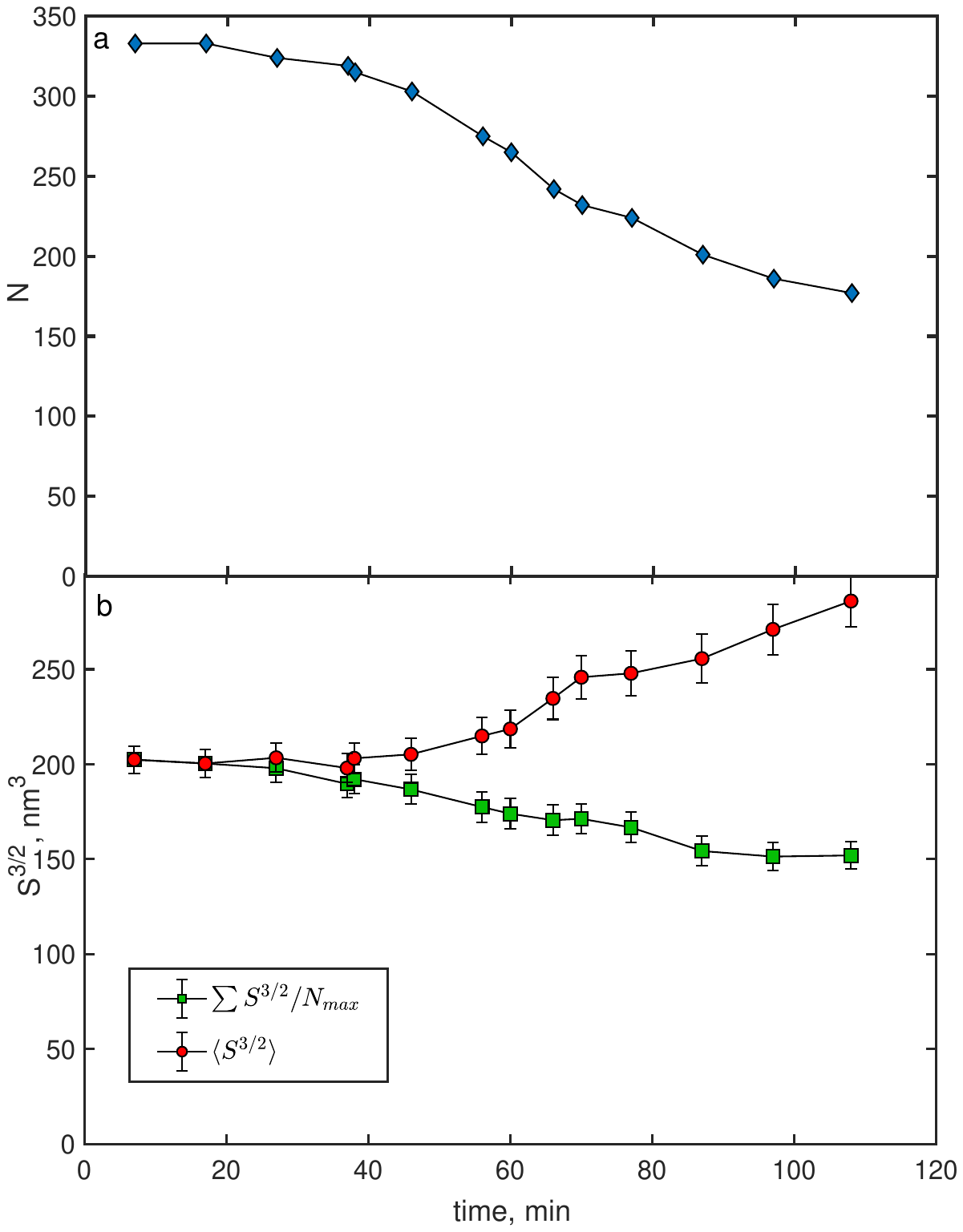}
\caption{\textbf{Ensemble evolution during annealing.}
  \textbf{(a)} Total NP count $N(t)$ versus time $t$ shows a monotonic decrease due to coalescence and evaporation.
  \textbf{(b)} Normalized total volume proxy $\sum_i S_i^{3/2}(t)/N_{\max}$ (green squares) and ensemble average volume proxy per particle, $\langle S^{3/2}\rangle(t)=N^{-1}(t)\sum_i S_i^{3/2}(t)$ (red circles). While $N(t)$ decreases, $\langle S^{3/2}\rangle$ increases, which is consistent with net coarsening.  The total volume proxy, $\sum_i S_i^{3/2}$ shows a gradual decrease  attributable to sublimation.
}\label{fig:evol}
\end{figure}
We quantify size using the footprint area $S_i(t)$ of each particle, that is expected to be related to its volume as
\begin{equation}
V_i(t)=f_i(t)\,S_i^{3/2}(t),
\end{equation}
where $f_i(t)$ is a shape factor. If the overall shape and contact angle remain approximately constant, $f_i(t)$ is effectively time independent. Figure~\ref{fig:merge} supports this assumption: for three NPs undergoing two consecutive mergers, the sum of areas $S_1+S_2+S_3$ is not conserved, whereas $S_1^{3/2}+S_2^{3/2}+S_3^{3/2}$ is nearly constant, which is consistent with volume conservation. This allows us to use $S_i^{3/2}$ as NP volume proxy, and track the ensemble volume up to a single scale factor:
\begin{equation}
V(t)=f\,\sum_i S_i^{3/2}(t).
\end{equation}
For a near-spherical cap with contact angle $\theta>90^\circ$,
\(
f=\frac{2(1-\cos\theta)}{3\sqrt{\pi}}
\);
for $\theta\approx 115^\circ$ (Ref.~\cite{stach2021}), $f\approx 0.5$.

Figure~\ref{fig:evol} summarizes the ensemble evolution within the analyzed field of view ($1000\,\mathrm{px}\times1000\,\mathrm{px}$, that corresponds to  $33630~\mathrm{nm}^2$). The total particle count $N(t)$ decreases steadily, the normalized sum $\sum_i S_i^{3/2}(t)$ shows a slow decline, and the average $\langle S^{3/2}\rangle(t)=N^{-1}(t)\sum_i S_i^{3/2}(t)$ increases with time. At first glance this resembles classical coarsening; however, the simultaneous loss of total volume indicates that mass is being removed from the field of view, inconsistent with Ostwald ripening as the dominant mechanism.

To disentangle the pathways, we examine the size-change rate at the single-particle level (Fig.~\ref{fig:scatter})a. Using $\Delta S^{3/2}$ over a fixed interval $\Delta t=10$~min as a proxy for growth, the data split into two distinct clusters. One cluster exhibits large positive jumps, reflecting coalescence events driven by lateral motion and collisions. This interpretation is supported by a strong temporal  correlation between the number of such events with the decrease of total NP number, demonstrated in Fig.~\ref{fig:scatter})b.  The second group of events forms a narrow band at small negative rates ($\Delta S^{3/2}/\Delta t\!\approx\!-0.8~\mathrm{nm}^3/\mathrm{min}$), which we attribute to sublimation. Two features are noteworthy: (i) coalescence is largely responsible for the drop in $N(t)$, while sublimation produces the gradual loss of total volume; and (ii) the net sublimation rate is nearly independent of particle size and shows surprisingly strong particle-to-particle variability even among similarly sized NPs. Such variability has previously been observed in refs. \cite{stach2021,eric2023}. Below we give it a quantitative description, and build a  unified theoretical model that gives an explanation top both effects. 

The weak size dependence runs counter to the intuition from Gibbs-Thomson-driven Ostwald ripening, in which curvature should strongly modulate evaporation/condensation rates. Below we demonstrate that a substrate-mediated evaporation naturally explains both observations: an approximately flat (size-independent) mean shrinkage profile and the broad dispersion of apparent rates.

\begin{figure}[h!]
\centering   
\includegraphics[width=1\linewidth]{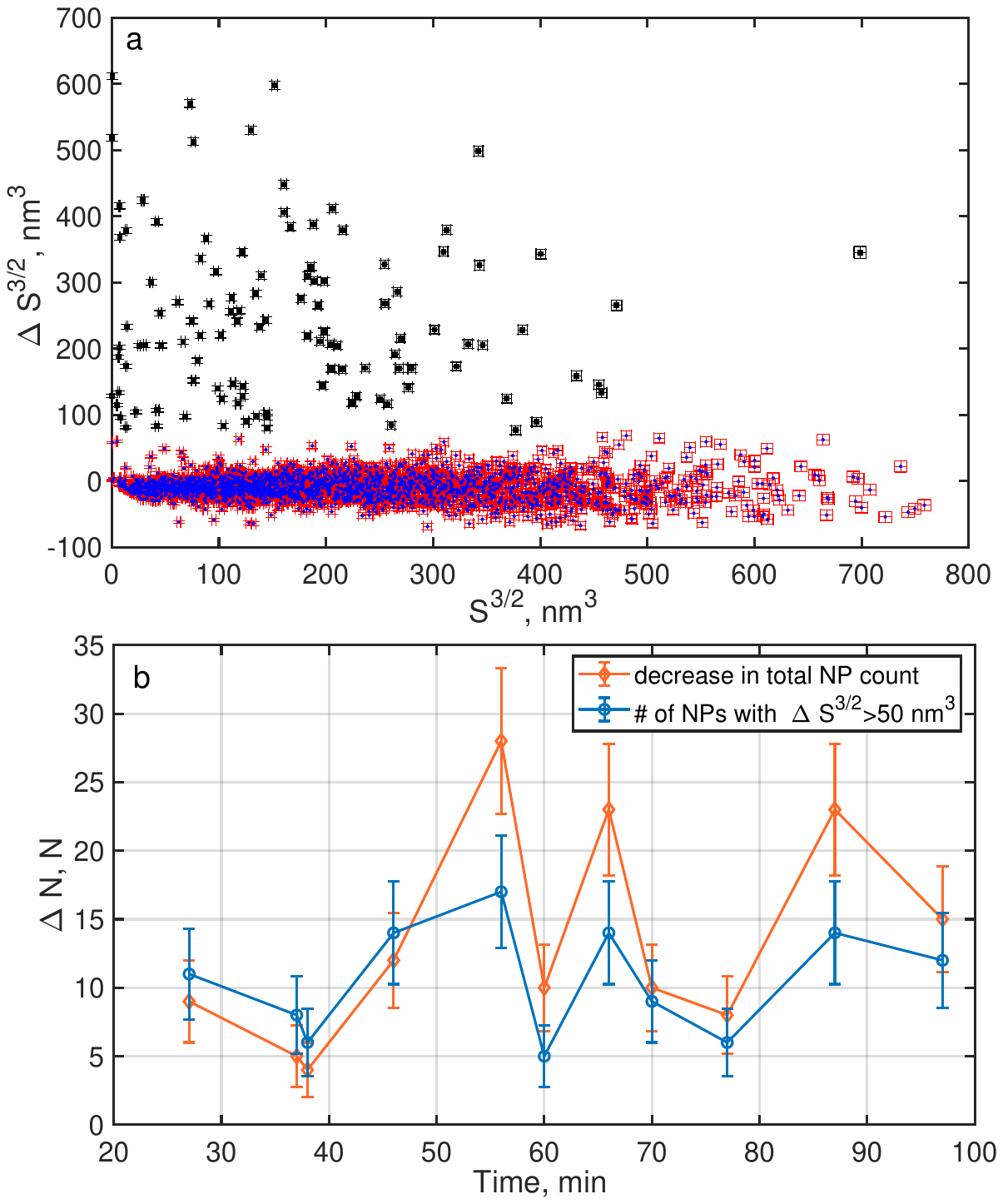}
\caption{\textbf{Bimodal distribution of NP size increments.}
  a) Scatter plot of the volume proxy change, $\Delta S^{3/2}$, versus $S^{3/2}$. Each point corresponds to a NP size change over a time interval $\Delta t=10$~min. Two distinct clusters emerge: large positive $\Delta S^{3/2}$ values indicate coalescence events (black crosses), while a narrow band in the vicinity of zero reflects sublimation-dominated size change (red dots). The solid line shows the mean sublimation rate of binned data as a function of $S^{3/2}$, with whiskers representing the standard deviation within each bin. b) Correlation between the number of NPs exhibiting large size increase  ($\Delta S^{3/2}>50 nm^3$) (blue circles), and the decrease of the \rev{ total particle count (orange diamonds). Both changes are taken between the two consecutive measurements. } This supports our interpretation of large size increases as coalescence events. 
 }\label{fig:scatter}
\end{figure}

\begin{figure}[h!]
\centering   
\includegraphics[width=1\linewidth]{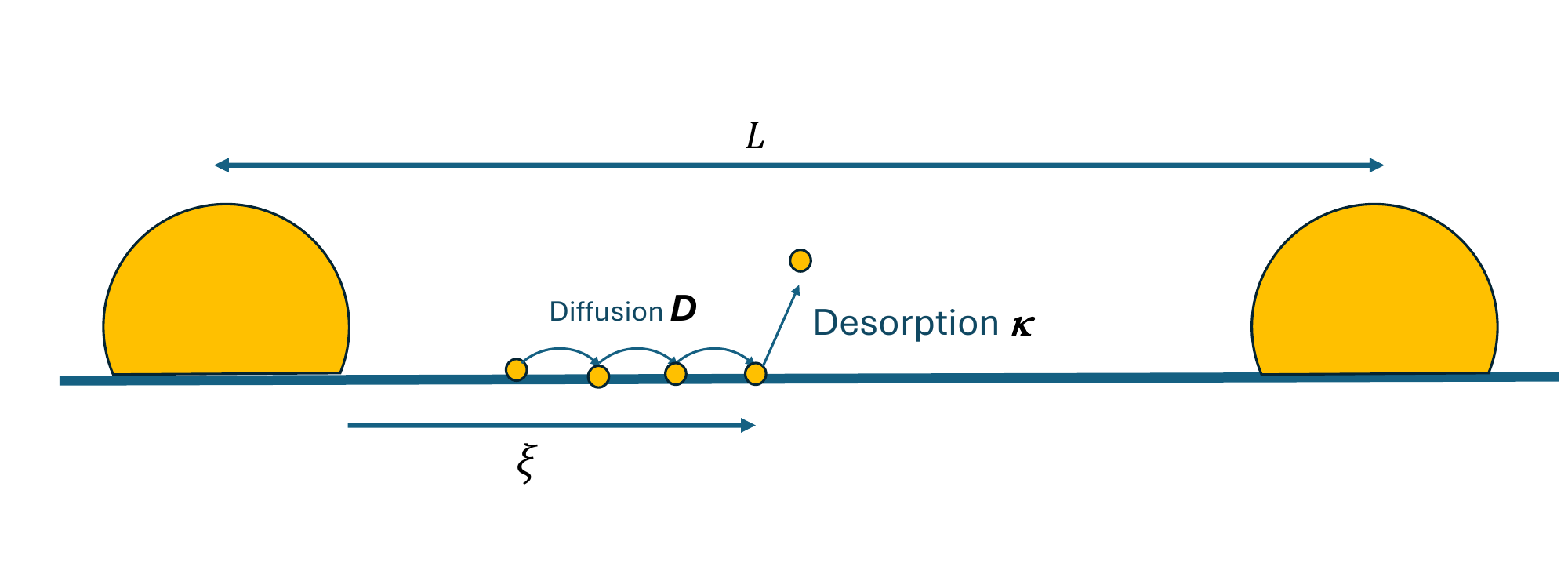}
\caption{\textbf{Schematic of substrate-mediated evaporation.}
  Supported NP of radius $R$ on a substrate with adatom diffusion length $\xi$ and interparticle spacing $L$. When $L\gtrsim \xi$, particles behave quasi-independently; evaporation includes (i) direct sublimation from the NP surface and (ii) desorption of adatoms generated within a substrate ring of order $\xi$ around each NP. The latter yields an approximately size-independent contribution to the per-particle mass loss.
}\label{fig:model}
\end{figure}

\subsection{Sublimation kinetics}
Sublimation of supported NPs involves two coupled pathways: (i) direct atom loss from the NP surface and (ii) atom departure from NP to the substrate followed by its 2D diffusion and desorption. Away from the NPs, the dynamics of adatom concentration $C(\mathbf{r})$ follows the following equation:
\begin{equation}
\dot C= D_a \nabla^2 C - Q C
\end{equation}
Here $D_a$ is the adatom surface diffusion coefficient and $Q$ is the desorption rate. The associated \rev{(bare)} screening length 
\begin{equation}
\rev{\xi_0}=\sqrt{\frac{D_a}{Q}}
\end{equation}
is the typical range over which adatoms explore the substrate before desorption (see Fig.~\ref{fig:model}) \cite{stach2021}. 
A simple Arrhenius estimate gives $Q\simeq \nu_0e^{-E_a/k_BT}$ and $D_a \simeq \frac{\nu_0a^2}{4}e^{-E_d/k_BT}$, with $E_a$ the adsorption energy on the substrate, $E_d$ the diffusion barrier, $\nu_0$ the attempt frequency, and $a$ an atomic length scale. According to ref.~\cite{eric2023}, the diffusion barrier is very low and $E_a\approx 1$~eV. Assuming $a\simeq 1~\text{\AA}$, we obtain the estimate
\begin{equation}
\rev {\xi_0}\simeq a e^{E_a/2k_BT}\sim 5~\text{nm}.
\label{eq:xi}
\end{equation}

\begin{figure}[th!]
\centering   
\includegraphics[width=\linewidth]{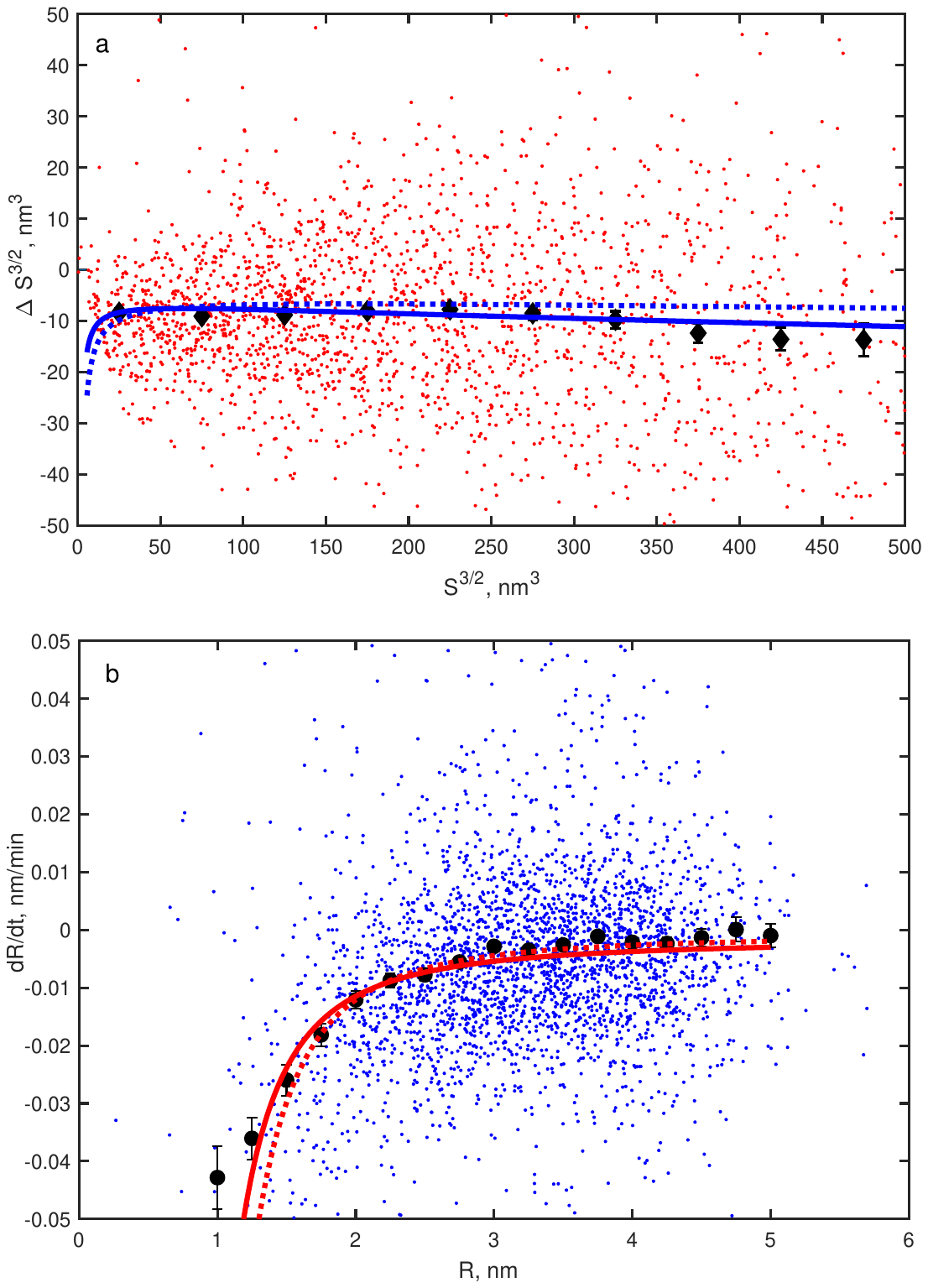}
\caption{\textbf{Sublimation regime and model comparison.} \textbf{(a)} Zoom of the sublimation-dominated band from Fig.~\ref{fig:scatter} with binned average rates represented by black data points. \rev{Error bars for  individual data points  (shown in Fig.~\ref{fig:scatter}) are ommitted for clarity.} The model fits are based on Eq.~\eqref{eq:v_dot}, with  \rev{$\xi_0=2.0$ nm (solid),  and  $\xi_0=4.0$ nm (dotted)}. Both fits assume $R_\gamma=3$~nm. \rev{Additional theoretical curves for various values of $\xi_0$ are presented in SI.}  \textbf{(b)} The same experimental data and model fits as in (a) represented as   radial rate $dR/dt$ versus $R$.  }
\label{fig:scatter2}
\end{figure}

The sublimation regime is determined by how $\xi$ compares to the particle radius $R=\sqrt{\langle S\rangle/\pi}\simeq 2$~nm (rms) and interparticle spacing $L\simeq 10$~nm. When $L\gtrsim \xi$, material exchange between particles is weak and each NP behaves quasi-independently.  In this limit, curvature (Gibbs-Thomson) corrections modulate the local equilibrium but do not produce classical Ostwald ripening (i.e., no net mass transfer from small to large particles). \rev{ The coarsening behavior is only  expected once the bare screening length $\xi_0$ significantly exceeds the typical interparticle distance. }

\begin{rev_par}

At the particle edge, $|\mathbf x-\mathbf x_i|=r_i\sin \theta$, the local adatom concentration is assumed to be in equilibrium with the NP,
\begin{equation}
C_{\mathrm{eq}}(R_i)=\frac{U}{Qv_a}\exp\!\left(\frac{R_\gamma}{R_i}\right),
\label{eq:Ceq_main}
\end{equation}
where \(U\) has units of \([\mathrm{volume}/(\mathrm{area}\cdot \mathrm{time})]\) and therefore has the physical meaning of an effective evaporation velocity from a flat interface. The capillary length is \(R_\gamma=2\gamma v_a/(k_BT)\), where \(\gamma\) is the surface energy and \(v_a\) is the atomic volume. While the surface energy of small Au NPs may deviate from the bulk value, recent studies \cite{Holec2020} indicate that it remains nearly constant for particle radii above \(\approx 3\) nm. Using \(\gamma\approx 1.5~\mathrm{J/m^2}\) for solid Au gives \(R_\gamma\approx 3\) nm \cite{tyson1977surface}.

To describe the sublimation kinetics, we developed a two-scale self-consistent theory (see Supporting Information) in which the substrate adatom concentration \(C(\mathbf x)\) is decomposed into (i) a localized near-field distortion around a tagged NP and (ii) a coarse-grained background generated by the rest of the ensemble. 
On coarse scales, each NP is represented as a localized drain/source center at \(\mathbf x_i\). The coarse-grained steady state equation for \(C(\mathbf x)\) is
\begin{align}
\sum_i 2\pi \kappa_i D_a \bigl[C(\mathbf x_i)-&C_{\mathrm{eq}}(R_i)\bigr] \delta(\mathbf x-\mathbf x_i)= \nonumber
\\
&=D_a\nabla^2 C-Q C \label{eq:cg_delta}
\end{align}
Here  \(\kappa_i\) is a dimensionless geometric kernel controlling substrate-mediated exchange for particle \(i\) (defined below).  Equation~\eqref{eq:cg_delta} implies  that neighboring particles provide an additional collective sink for adatoms and therefore renormalize the screening length from its bare value \(\xi_0=\sqrt{D_a/Q}\) to
\begin{equation}
\xi=\sqrt{\frac{D_a}{Q+2\pi D_a n\langle \kappa\rangle}}
=\frac{\xi_0}{\sqrt{1+2\pi n\xi_0^2\langle \kappa\rangle }}
\label{eq:xi_sc_main}
\end{equation}
Here  \(n=\frac{4}{\pi L^2}\) is the 2D density of NPs. This result establishes a crossover between the single-particle problem with substrate-mediated evaporation and collective Ostwald ripening behavior in 2D in the absence of evaporation \cite{Ostwald2D}. In that limit, the evaporation-based bare screening length $\xi_0$ is replaced with a new length $1/\sqrt{2\pi n\langle \kappa \rangle }$, which is comparable to the inter-$NP$  distance $L$.  

For a spatially uniform steady state, Eq.~\eqref{eq:cg_delta} yields the mean-field background
\begin{equation}
C^*=
2\pi \xi^2 n \langle \kappa\, C_{\mathrm{eq}}(R)\rangle
\label{eq:Cstar_main}
\end{equation}

Having determined \(C^*\) and \(\xi\) from the coarse-grained theory, the near-field concentration profile around a tagged particle \(i\) at \(\mathbf x_i\) follows from the screened diffusion equation. Writing \(r\equiv|\mathbf x-\mathbf x_i|\) and decomposing the local field as \(C(r)=C^*+c(r)\) results in 
\begin{equation}
c(r)=\bigl[C_{\mathrm{eq}}(R_i)-C^*\bigr]\frac{K_0(r/\xi)}{K_0(x_i)},
\quad x_i=\frac{R_i\sin\theta}{\xi}
\label{eq:Cprofile_main}
\end{equation}
Here and below $K$ is modified Bessel function of the  second kind.  Calculating the edge flux, one obtains the geometric kernel
\begin{equation}
\kappa_i
=
x_i\frac{K_1(x_i)}{K_0(x_i)}
\label{eq:kappa_main}
\end{equation}
Equations~\eqref{eq:xi_sc_main} and \eqref{eq:kappa_main} determine the effective screening length \(\xi\) self-consistently through the ensemble average \(\langle\kappa\rangle\).

By combining the substrate-mediated exchange rate implied by Eq.~\eqref{eq:Cprofile_main} with direct evaporation from the exposed NP surface, we obtain the mean volume shrinkage rate for NPs of size \(R_i\):
\begin{align}
\overline{\dot V}_{R_i}
=&
-2\pi U\xi^2
\Bigg(
\chi_i(1-\cos\theta)\left(\frac{R_i}{\xi}\right)^2+
\label{eq:v_dot}\\
&+
\kappa_i
\left[
\chi_i
+
2\pi n\xi_0^2\left(\langle\kappa\rangle\chi_i-\langle\kappa\chi\rangle
\right)\right]
\Bigg) \nonumber
\end{align}
where \(\chi_i=\exp(R_\gamma/R_i)\).  Our theory  provides a unified description that combines effects of both direct and substrate-mediated evaporation with 2D Ostwald ripening, thus generalizing the classical result by Yao et al \cite{Ostwald2D}.   The first term inj \eqref{eq:v_dot} describes direct evaporation from the exposed NP surface, the second is the local substrate-mediated loss term, and the last one is a collective Ostwald-like correction arising from the self-consistent background concentration generated by the particle ensemble. This correction vanishes for a monodisperse population and is therefore a direct consequence of polydispersity.
Equation~\eqref{eq:v_dot} reduces to the isolated-particle result when the distance between  NPs significantly exceeds the bare screening length $\xi_0$ , i.e.,  \(2\pi n\xi_0^2\ll 1\) or when the distribution is sufficiently narrow that \(\langle\kappa\rangle\chi_i-\langle\kappa\chi\rangle\approx 0\). The theory naturally predicts a crossover between nearly independent sublimation and weak collective coupling through the substrate.
 \end{rev_par}

 For  $R_i\lesssim \xi \lesssim L$, second term in Eq.~\eqref{eq:v_dot} \rev{becomes dominant. As a result,}   the evaporation rate per particle approaches a nearly constant value, in agreement with the experimentally observed flat $dV/dt$ profile (Fig.~\ref{fig:scatter2}). \rev{The quantitative agreement is indeed very good and robust with respect to model parameters.  While a somewhat better fit is achieved for  $\xi_0=2.0$ nm,  the agreement remains reasonable for larger values of bare screening length,  $\xi_0=4.0$nm, which is more consistent with  our simple estimate,  Eq.~\eqref{eq:xi}.  Variations in capillary length, $R_\gamma$ within reasonable limits ($1 - 3$ nm)} affect only the smallest particles ($R\sim 1$~nm), which are underrepresented in the data.

From fitting, we obtain an effective evaporation speed $U \approx 1.2\times 10^{-4}$~nm\,s$^{-1}$ (for $f\approx 0.5$ and $\xi=5.0$~nm).  This is a surprisingly low rate for the experimental conditions.  Indeed, one can use the classical Hertz-Knudsen equation to estimate the sublimation rate per NP to be 
\begin{equation}
   U_{HK}=\frac {\alpha P(T)}{\rho} \sqrt{\frac{M}{2\pi RT}}
\end{equation}
Here  $M$ and $\rho$ are the molar mass and density of gold. \rev{This result is based on the expected collision rate of gold aloms at a saturated vapor pressure $P(T)\approx  10^{-4} Pa$ \cite{vapour}, corrected  by the sticking coefficient $\alpha \sim 1$. } Once we plug in all the parameters we arrive at the theoretical prediction $U_{HK}\simeq 3 $~ nm/sec.  As one can see, the observed sublimation rate is several orders of magnitude smaller than expected.   A plausible explanation for this anomalously slow sublimation could be the formation of Au-Si eutectic, possibly facilitated by the electron beam \cite{Panciera2019,AuSi2022_Eutectic_Dynamics}. However, this evidence is indirect, and the phenomenon requires further in-depth studies.

\begin{figure}[h!]
\centering   
\includegraphics[width=\linewidth]{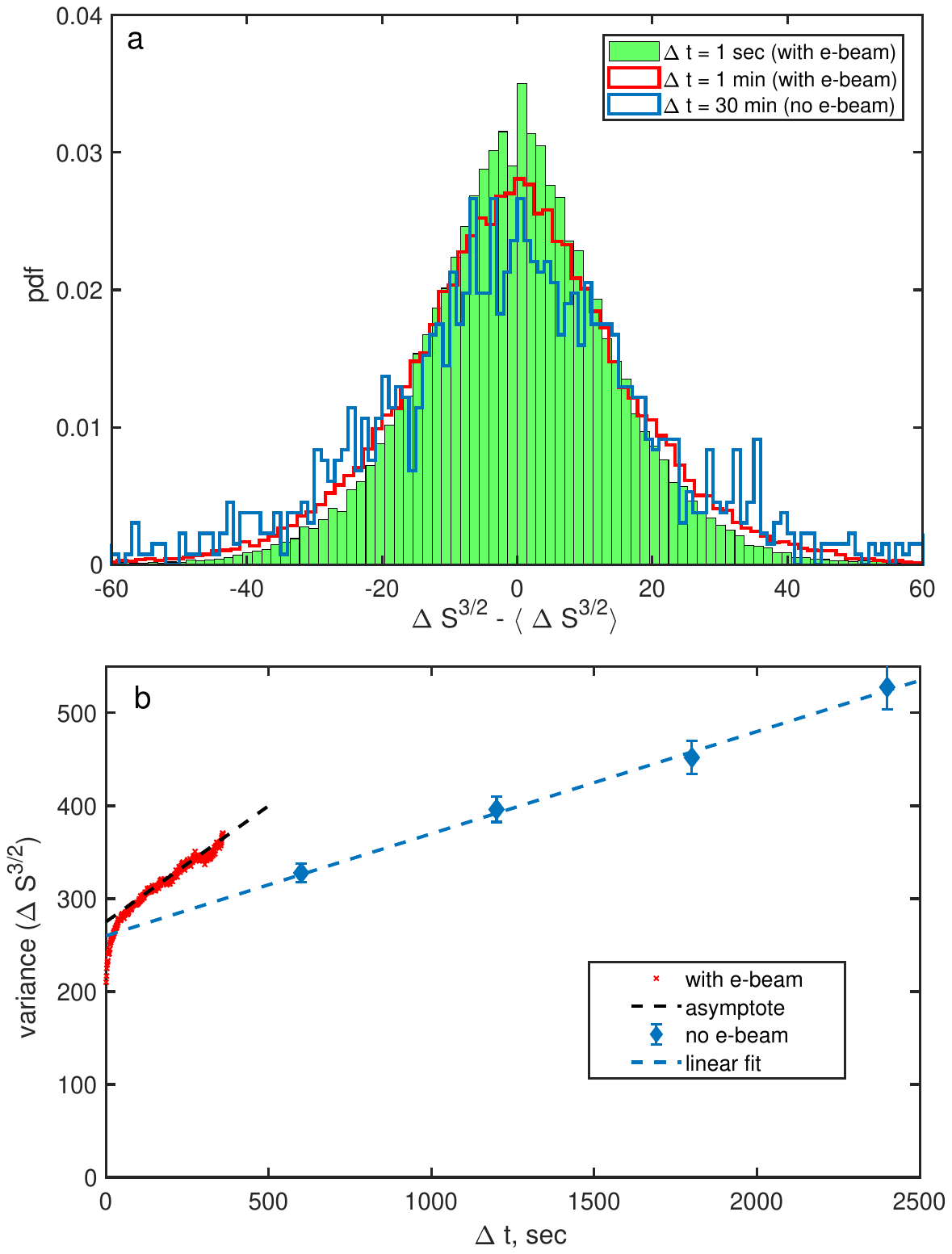}
\caption{\textbf{Fluctuation statistics of volume proxy.}
  \textbf{(a)} Histograms of $\Delta S^{3/2}$ for $\Delta t=1$~s and 1~min \rev{(in the presence of e-beam) and $\Delta t=30$~min (no e-beam)}. Distributions are similar, with a slightly narrower width at $\Delta t=1$~s, indicating dominant rapid shape fluctuations with a relaxation time $\lesssim 1$~min (with beam).
  \textbf{(b)} Variance of $\Delta S^{3/2}$ versus $\Delta t$. Red line: the early data taken under continuous e-beam exposure, with linear asymptotic behavior (dashed black) reached after a transient period of the order of 1 min.  Diamonds: the longer term evolution with e-beam off. It also exhibits a linear behavior (blue dashed), consistent with a random walk in the volume space.
}\label{fig:dV}
\end{figure}

\subsection{Stochastic effects and NP diffusion}

A striking feature in Figs.~\ref{fig:scatter}--\ref{fig:scatter2} is the wide spread of apparent growth rates even among NPs of similar size. Recall that we track size via the volume proxy $S^{3/2}$ with $V_i=f_i(t)\,S_i^{3/2}(t)$; short-time changes therefore reflect not only true mass exchange but also rapid shape fluctuations through $f_i(t)$.

To separate these contributions, we analyze the statistics of $\Delta S^{3/2}$ over different time increments (Fig.~\ref{fig:dV}). The histograms in Fig.~\ref{fig:dV}a for $\Delta t=1$~s and $1$~min (beam on) and $\Delta t=10$~min (beam off) are very similar, with only a slight narrowing at $\Delta t=1$~s. This indicates that on sub-minute timescales the variance is dominated by fast shape relaxations rather than persistent mass change; the corresponding shape-relaxation time is $\lesssim 1$~min under our imaging conditions. This shape fluctuations are fully consistent with surface rearrangements directly observed and modeled in ref. \cite{eric2023}.  

At longer times, the variance of $\Delta S^{3/2}$ grows linearly with $\Delta t$ (Fig.~\ref{fig:dV}b), revealing a genuine stochastic evolution of particle volume that accumulates like a random walk \cite{chandler1987}. We model the dynamics with a minimal Langevin equation in volume space,
\begin{equation}
  \dot V = \overline{\dot V}_{R_i} + \eta_i(t),
\end{equation}
where $\overline{\dot V}_{R_i}$ is the mean drift from substrate-mediated evaporation (Eq.~\ref{eq:v_dot}) and $\eta_i(t)$ is a short-correlated noise term with
\(
\langle \eta_i(t)\eta_i(t') \rangle = \Lambda\,\delta(t-t').
\)
Here $\Lambda/2$ plays the role of an effective diffusion coefficient in volume space and sets the slope of the long-time variance:
\rev{\begin{equation}
    \textrm{var} \left(V(t+\Delta t)-V(t)\right)=f^2  \textrm{var}\left(\Delta S^{3/2}\right)=\Lambda \Delta t
\end{equation}}
From Fig.~\ref{fig:dV}b (using $f\approx 0.5$ to convert $S^{3/2}$ to $V$), we obtain
\begin{align}
&\Lambda \approx 0.06~\mathrm{nm^6\,s^{-1}} &\quad \text{(beam on)}\\
&\Lambda \approx 0.03~\mathrm{nm^6\,s^{-1}} &\quad \text{(beam off)}
\end{align}

The magnitude of $\Lambda$ is consistent with a simple scaling estimate in which random attachment/detachment events of adatoms provide steps of size $\sim v_a$ in $V$, occurring at a rate $\sim 4\pi R D_a C_{\rm eq}(R)/a$ for a particle of radius $R$ (here $a$ an atomic length). One can connect this estimate to the parameters of our earlier evaporation model, in particular $D_a = \xi_0^2 Q$ based on  Eq. (\ref{eq:xi}),  and $Q C=U/v_a$ is the evaporation rate from unit substrate area, consistent with the observed evaporation speed $U$. Together, this gives   
\begin{align}
\Lambda \simeq &\left\langle \frac{4\pi  R v^2_a}{a} D_a C_{\rm eq}(R) \right \rangle \sim \\
&\sim 4\pi \xi_0^2 U  v_a^{2/3} \langle R e^{R_\gamma/R} \rangle \sim   0.01\frac{nm^6}{s} \nonumber
\end{align}
This estimate is within an order of magnitude of the measured values. In this analysis, we have assumed $\xi_0 = 5~\mathrm{nm}$, consistent with the earlier theoretical estimate in Eq.~\eqref{eq:v_dot}, and with the fits of experimental data in Figure \ref{fig:scatter2}.  This support the interpretation that the observed noise originates primarily from intermittent adatom exchange, with an additional contribution from rapid shape fluctuations. This picture is also consistent with results of Ref. \cite{eric2023} : the strong  surface rearrangements are linked to a very low kinetic barrier for atom detachment from NPs, and the observed strong diffusive drift of NP volume.     

\begin{figure}[h!]
\centering   
\includegraphics[width=1.1\linewidth]{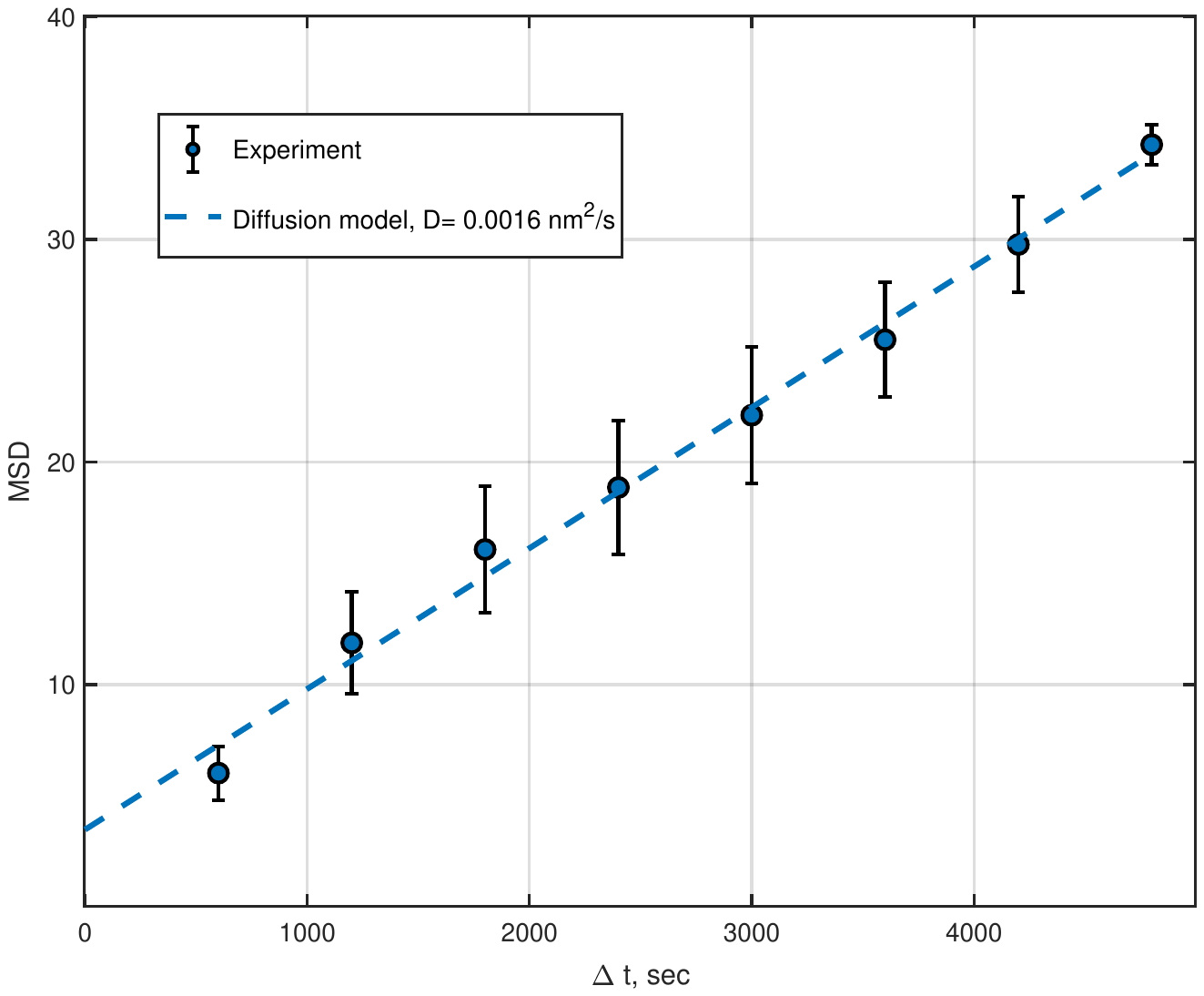}
\caption{\textbf{Lateral diffusion of nanoparticles.}
  Mean-squared displacement (MSD) of NP centroids versus time shows a linear dependence, consistent with Brownian-like surface diffusion. The extracted diffusion coefficient is $D_{\mathrm{np}}\approx 1.6\times 10^{-3}$~nm$^2$\,s$^{-1}$. 
}\label{fig:diff}
\end{figure}

A second stochastic phenomenon is the lateral motion of the NPs themselves. The mean-squared displacement of NP centroids grows linearly with time (Fig.~\ref{fig:diff}), consistent with diffusive motion \cite{chandler1987},
\begin{equation}
\mathrm{MSD}(\Delta t)= 4 D_{np} \Delta t,
\end{equation}
with an extracted diffusion coefficient
\(
D_{\mathrm{np}} = (1.6 \pm 0.1)\times 10^{-3}~\mathrm{nm^2\,s^{-1}}.
\)
This lateral diffusion sets the collision frequency and thus the rate of coalescence events observed in Fig.~\ref{fig:scatter}.

\section{Conclusions}

By combining \textit{in situ} TEM, particle-resolved tracking, and minimal modeling, we disentangled the average and stochastic contributions to the evolution of supported Au nanoparticles. \rev{A central outcome of this work is the development of a self-consistent theoretical framework that unifies substrate-mediated evaporation with collective 2D Ostwald-type coupling through the shared adatom field, while also incorporating stochastic single-particle fluctuations within the same description.} Surprisingly, the overall NP sublimation is orders of magnitude slower than a Hertz--Knudsen estimate. A plausible explanation of this discrepancy could be beam-assisted Au--Si eutectic formation. However, further dedicated studies are needed to establish the actual microscopic mechanism.

Due to the slow yet non-zero net evaporation, the total particle volume is not conserved, which suppresses classical Ostwald ripening. While the very small NPs ($\sim 1$ nm) do evaporate significantly faster, the lost material is not effectively transferred to larger particles. Instead, the average mass loss per NP exhibits a flat, nearly size-independent evaporation profile. \rev{Within our theory, this regime corresponds to $R \lesssim \xi \lesssim L$, where substrate-mediated desorption dominates over curvature-driven redistribution, while collective coupling through the renormalized screening length remains present but weak.}

On top of this mean drift, individual NPs display substantial particle-to-particle variability: their apparent growth or shrinkage rates fluctuate strongly even at similar sizes, and their volume trajectories accumulate in a random-walk-like manner. A minimal description captures these observations: a substrate-mediated evaporation term accounts for the nearly flat mean shrinkage profile, while a stochastic contribution associated with intermittent adatom attachment and detachment reproduces the measured fluctuation spectrum. \rev{The extracted noise strength is quantitatively consistent with microscopic adatom exchange rates, thereby linking trajectory-level statistics directly to the underlying atomistic processes.} In addition, the lateral motion of particle centroids is described by normal diffusion, with diffusion coefficient $D\approx 0.0016~\mathrm{nm^2/s}$. This diffusive motion sets the collision frequency and is responsible for the slow coalescence of NPs.

\begin{rev_par}
Importantly, the stochasticity uncovered here is of a different character from that emphasized in traditional population-balance descriptions of nanoparticle evolution. In such models, randomness typically enters at the ensemble level through discrete collision and coalescence events between particles. By contrast, our experiments reveal stochasticity at a finer scale: even in between collisions, individual particles undergo substantial fluctuations due to the discrete exchange of atoms with the substrate adatom reservoir. Thus, stochasticity is not merely a consequence of rare particle-particle merger events, but an intrinsic part of the continuous particle-substrate exchange dynamics.

Our study focuses on early and intermediate stages of supported nanoparticle evolution, prior to the onset of asymptotic self-similar coarsening typically targeted by long-time population-balance theories. By explicitly resolving single-particle trajectories, we show that deterministic drift, collective substrate-mediated coupling, and intrinsic stochasticity can all be comparable in magnitude under experimentally relevant conditions. In this sense, the present work provides microscopic expressions for both the mean drift and the fluctuation terms that could serve as physically grounded inputs for future ensemble-level descriptions of supported nanoparticle systems.

More broadly, this work demonstrates that the evolution of supported nanoparticles cannot be fully understood within either purely deterministic mean-field pictures or purely stochastic coarse-grained descriptions taken in isolation. Instead, the relevant physics emerges from their interplay: substrate-mediated transport sets the collective drift, while discrete atomic events generate measurable trajectory-level noise. By quantitatively linking these scales within a single framework, we provide a pathway toward predictive, experimentally grounded modeling of nanoparticle stability beyond asymptotic coarsening limits.
\end{rev_par}

Practically, our perspective elevates stochasticity from a nuisance to a design parameter. Incorporating it alongside average kinetics is essential for predictive control of nanoparticle stability across catalysis, plasmonics, and sensing.

\section*{Acknowledgements}
This research was conducted at  Brookhaven National Laboratory under Contract No. DE-SC0012704, where it used the Electron Microscopy, Theory and Computation facilities of the Center for Functional Nanomaterials (CFN), which is a U.S. Department of Energy Office of Science User Facility. J.P.H. and E.A.S. acknoweldge support by the National Science Foundation, Division of Materials Research's Metals and Metallic Nanostructures program, DMR-1809398 and DMR-2303084. 
\bibliography{NP_supp}

\providecommand{\latin}[1]{#1}
\makeatletter
\providecommand{\doi}
  {\begingroup\let\do\@makeother\dospecials
  \catcode`\{=1 \catcode`\}=2 \doi@aux}
\providecommand{\doi@aux}[1]{\endgroup\texttt{#1}}
\makeatother
\providecommand*\mcitethebibliography{\thebibliography}
\csname @ifundefined\endcsname{endmcitethebibliography}  {\let\endmcitethebibliography\endthebibliography}{}
\begin{mcitethebibliography}{38}
\providecommand*\natexlab[1]{#1}
\providecommand*\mciteSetBstSublistMode[1]{}
\providecommand*\mciteSetBstMaxWidthForm[2]{}
\providecommand*\mciteBstWouldAddEndPuncttrue
  {\def\EndOfBibitem{\unskip.}}
\providecommand*\mciteBstWouldAddEndPunctfalse
  {\let\EndOfBibitem\relax}
\providecommand*\mciteSetBstMidEndSepPunct[3]{}
\providecommand*\mciteSetBstSublistLabelBeginEnd[3]{}
\providecommand*\EndOfBibitem{}
\mciteSetBstSublistMode{f}
\mciteSetBstMaxWidthForm{subitem}{(\alph{mcitesubitemcount})}
\mciteSetBstSublistLabelBeginEnd
  {\mcitemaxwidthsubitemform\space}
  {\relax}
  {\relax}

\bibitem[Sápi \latin{et~al.}(2021)Sápi, Rajkumar, Kiss, Kukovecz, Kónya, and Somorjai]{Sapi2021_MetalNP_Catalysis}
Sápi,~A.; Rajkumar,~T.; Kiss,~J.; Kukovecz,~Ã.; Kónya,~Z.; Somorjai,~G.~A. Metallic Nanoparticles in Heterogeneous Catalysis. \emph{Catalysis Letters} \textbf{2021}, \emph{151}, 2153–2175\relax
\mciteBstWouldAddEndPuncttrue
\mciteSetBstMidEndSepPunct{\mcitedefaultmidpunct}
{\mcitedefaultendpunct}{\mcitedefaultseppunct}\relax
\EndOfBibitem
\bibitem[van Deelen \latin{et~al.}(2019)van Deelen, Hernández~Mejía, and de~Jong]{vanDeelen2019_MSI_NatCatal}
van Deelen,~T.~W.; Hernández~Mejía,~C.; de~Jong,~K.~P. Control of metal-support interactions in heterogeneous catalysts to enhance activity and selectivity. \emph{Nature Catalysis} \textbf{2019}, \emph{2}, 955–970\relax
\mciteBstWouldAddEndPuncttrue
\mciteSetBstMidEndSepPunct{\mcitedefaultmidpunct}
{\mcitedefaultendpunct}{\mcitedefaultseppunct}\relax
\EndOfBibitem
\bibitem[Kant \latin{et~al.}(2024)Kant, Beeram, Cao, dos Santos, González-Cabaleiro, García-Lojo, Guo, Joung, Kothadiya, Lafuente, Leong, Liu, Liu, Moram, Mahasivam, Maniappan, Quesada-González, Raj, Weerathunge, Xia, Yu, Abalde-Cela, Alvarez-Puebla, Bardhan, Bansal, Choo, Coelho, de~Almeida, Gómez-Graña, Grzelczak, Herves, Kumar, Lohmueller, Merkoçi, Montaño-Priede, Ling, Mallada, Pérez-Juste, Pina, Singamaneni, Soma, Sun, Tian, Wang, Polavarapu, and Santos]{NanoscaleHorizons2024_PlasmonicSensors}
Kant,~K. \latin{et~al.}  Plasmonic nanoparticle sensors: current progress{,} challenges{,} and future prospects. \emph{Nanoscale Horiz.} \textbf{2024}, \emph{9}, 2085--2166\relax
\mciteBstWouldAddEndPuncttrue
\mciteSetBstMidEndSepPunct{\mcitedefaultmidpunct}
{\mcitedefaultendpunct}{\mcitedefaultseppunct}\relax
\EndOfBibitem
\bibitem[Kim \latin{et~al.}(2021)Kim, Park, Jung, Yeom, and Yoo]{Sensors2021_LSPR_Review}
Kim,~D.~M.; Park,~J.~S.; Jung,~S.-W.; Yeom,~J.; Yoo,~S.~M. Biosensing Applications Using Nanostructure-Based Localized Surface Plasmon Resonance Sensors. \emph{Sensors} \textbf{2021}, \emph{21}, 3191\relax
\mciteBstWouldAddEndPuncttrue
\mciteSetBstMidEndSepPunct{\mcitedefaultmidpunct}
{\mcitedefaultendpunct}{\mcitedefaultseppunct}\relax
\EndOfBibitem
\bibitem[Roldán~Cuenya and Ba{\~n}ares(2024)Roldán~Cuenya, and Ba{\~n}ares]{ChemRev2024_OperandoIntro}
Roldán~Cuenya,~B.; Ba{\~n}ares,~M.~A. Introduction: Operando and In Situ Studies in Catalysis and Electrocatalysis. \emph{Chemical Reviews} \textbf{2024}, \emph{124}, 8011--8013\relax
\mciteBstWouldAddEndPuncttrue
\mciteSetBstMidEndSepPunct{\mcitedefaultmidpunct}
{\mcitedefaultendpunct}{\mcitedefaultseppunct}\relax
\EndOfBibitem
\bibitem[Liu \latin{et~al.}(2024)Liu, Zhao, Han, Jia, Hong, and Liu]{NanoToday2024_InSituVideo_DL}
Liu,~S.; Zhao,~Q.; Han,~S.; Jia,~Z.; Hong,~X.; Liu,~W. Dynamics of catalyst nanoparticles quantified from in situ TEM video. \emph{Nano Today} \textbf{2024}, \emph{59}, 102505\relax
\mciteBstWouldAddEndPuncttrue
\mciteSetBstMidEndSepPunct{\mcitedefaultmidpunct}
{\mcitedefaultendpunct}{\mcitedefaultseppunct}\relax
\EndOfBibitem
\bibitem[Simonsen \latin{et~al.}(2011)Simonsen, Chorkendorff, Dahl, Skoglundh, Sehested, and Helveg]{SIMONSEN2011}
Simonsen,~S.~B.; Chorkendorff,~I.; Dahl,~S.; Skoglundh,~M.; Sehested,~J.; Helveg,~S. Ostwald ripening in a Pt/SiO2 model catalyst studied by in situ TEM. \emph{Journal of Catalysis} \textbf{2011}, \emph{281}, 147--155\relax
\mciteBstWouldAddEndPuncttrue
\mciteSetBstMidEndSepPunct{\mcitedefaultmidpunct}
{\mcitedefaultendpunct}{\mcitedefaultseppunct}\relax
\EndOfBibitem
\bibitem[Hansen \latin{et~al.}(2013)Hansen, DeLaRiva, Challa, and Datye]{Hansen2013}
Hansen,~T.~W.; DeLaRiva,~A.~T.; Challa,~S.~R.; Datye,~A.~K. Sintering of Catalytic Nanoparticles: Particle Migration or Ostwald Ripening? \emph{Accounts of Chemical Research} \textbf{2013}, \emph{46}, 1720–1730\relax
\mciteBstWouldAddEndPuncttrue
\mciteSetBstMidEndSepPunct{\mcitedefaultmidpunct}
{\mcitedefaultendpunct}{\mcitedefaultseppunct}\relax
\EndOfBibitem
\bibitem[DeLaRiva \latin{et~al.}(2013)DeLaRiva, Hansen, Challa, and Datye]{Delariva2013}
DeLaRiva,~A.~T.; Hansen,~T.~W.; Challa,~S.~R.; Datye,~A.~K. In situ Transmission Electron Microscopy of catalyst sintering. \emph{Journal of Catalysis} \textbf{2013}, \emph{308}, 291--305, 50th Anniversary Special Issue\relax
\mciteBstWouldAddEndPuncttrue
\mciteSetBstMidEndSepPunct{\mcitedefaultmidpunct}
{\mcitedefaultendpunct}{\mcitedefaultseppunct}\relax
\EndOfBibitem
\bibitem[Visser \latin{et~al.}(2023)Visser, Turner, Stewart, Vandegehuchte, van~der Hoeven, and de~Jongh]{Visser2023}
Visser,~N.~L.; Turner,~S.~J.; Stewart,~J.~A.; Vandegehuchte,~B.~D.; van~der Hoeven,~J. E.~S.; de~Jongh,~P.~E. Direct Observation of Ni Nanoparticle Growth in Carbon-Supported Nickel under Carbon Dioxide Hydrogenation Atmosphere. \emph{ACS Nano} \textbf{2023}, \emph{17}, 14963–14973\relax
\mciteBstWouldAddEndPuncttrue
\mciteSetBstMidEndSepPunct{\mcitedefaultmidpunct}
{\mcitedefaultendpunct}{\mcitedefaultseppunct}\relax
\EndOfBibitem
\bibitem[Woehl \latin{et~al.}(2012)Woehl, Evans, Arslan, Ristenpart, and Browning]{Woehl2012}
Woehl,~T.~J.; Evans,~J.~E.; Arslan,~I.; Ristenpart,~W.~D.; Browning,~N.~D. Direct in Situ Determination of the Mechanisms Controlling Nanoparticle Nucleation and Growth. \emph{ACS Nano} \textbf{2012}, \emph{6}, 8599–8610\relax
\mciteBstWouldAddEndPuncttrue
\mciteSetBstMidEndSepPunct{\mcitedefaultmidpunct}
{\mcitedefaultendpunct}{\mcitedefaultseppunct}\relax
\EndOfBibitem
\bibitem[Asoro \latin{et~al.}(2013)Asoro, Kovar, and Ferreira]{Asoro2013}
Asoro,~M.~A.; Kovar,~D.; Ferreira,~P.~J. In situ Transmission Electron Microscopy Observations of Sublimation in Silver Nanoparticles. \emph{ACS Nano} \textbf{2013}, \emph{7}, 7844–7852\relax
\mciteBstWouldAddEndPuncttrue
\mciteSetBstMidEndSepPunct{\mcitedefaultmidpunct}
{\mcitedefaultendpunct}{\mcitedefaultseppunct}\relax
\EndOfBibitem
\bibitem[Li \latin{et~al.}(2019)Li, Wang, Li, and Deepak]{Li2019}
Li,~J.; Wang,~Z.; Li,~Y.; Deepak,~F.~L. In Situ Atomic-Scale Observation of Kinetic Pathways of Sublimation in Silver Nanoparticles. \emph{Advanced Science} \textbf{2019}, \emph{6}, 1802131\relax
\mciteBstWouldAddEndPuncttrue
\mciteSetBstMidEndSepPunct{\mcitedefaultmidpunct}
{\mcitedefaultendpunct}{\mcitedefaultseppunct}\relax
\EndOfBibitem
\bibitem[Malyi and Rabkin(2012)Malyi, and Rabkin]{Malyi2012}
Malyi,~O.; Rabkin,~E. The effect of evaporation on size and shape evolution of faceted gold nanoparticles on sapphire. \emph{Acta Materialia} \textbf{2012}, \emph{60}, 261–268\relax
\mciteBstWouldAddEndPuncttrue
\mciteSetBstMidEndSepPunct{\mcitedefaultmidpunct}
{\mcitedefaultendpunct}{\mcitedefaultseppunct}\relax
\EndOfBibitem
\bibitem[Liu \latin{et~al.}(2021)Liu, Yuan, Wang, and Wang]{Materials2021_Sublimation_Melting}
Liu,~Y.; Yuan,~H.; Wang,~H.; Wang,~Z. In Situ Transmission Electron Microscopy Investigation of Melting/Evaporation Kinetics in Anisotropic Gold Nanoparticles. \emph{Materials} \textbf{2021}, \emph{14}, 7332\relax
\mciteBstWouldAddEndPuncttrue
\mciteSetBstMidEndSepPunct{\mcitedefaultmidpunct}
{\mcitedefaultendpunct}{\mcitedefaultseppunct}\relax
\EndOfBibitem
\bibitem[Meng \latin{et~al.}(2013)Meng, Yanagida, Kanai, Suzuki, Nagashima, Xu, Zhuge, Klamchuen, He, Rahong, Kai, and Kawai]{PRE2013_Au_Evaporation_Pressure}
Meng,~G.; Yanagida,~T.; Kanai,~M.; Suzuki,~M.; Nagashima,~K.; Xu,~B.; Zhuge,~F.; Klamchuen,~A.; He,~Y.; Rahong,~S.; Kai,~S.; Kawai,~T. Pressure-induced evaporation dynamics of gold nanoparticles on oxide substrate. \emph{Physical Review E} \textbf{2013}, \emph{87}, 012405\relax
\mciteBstWouldAddEndPuncttrue
\mciteSetBstMidEndSepPunct{\mcitedefaultmidpunct}
{\mcitedefaultendpunct}{\mcitedefaultseppunct}\relax
\EndOfBibitem
\bibitem[Horwath \latin{et~al.}(2021)Horwath, Voorhees, and Stach]{stach2021}
Horwath,~J.~P.; Voorhees,~P.~W.; Stach,~E.~A. Quantifying Competitive Degradation Processes in Supported Nanocatalyst Systems. \emph{Nano Letters} \textbf{2021}, \emph{21}, 5324--5329\relax
\mciteBstWouldAddEndPuncttrue
\mciteSetBstMidEndSepPunct{\mcitedefaultmidpunct}
{\mcitedefaultendpunct}{\mcitedefaultseppunct}\relax
\EndOfBibitem
\bibitem[Horwath \latin{et~al.}(2023)Horwath, Lehman-Chong, Vojvodic, and Stach]{eric2023}
Horwath,~J.~P.; Lehman-Chong,~C.; Vojvodic,~A.; Stach,~E.~A. Surface Rearrangement and Sublimation Kinetics of Supported Gold Nanoparticle Catalysts. \emph{ACS Nano} \textbf{2023}, \emph{17}, 8098--8107, PMID: 37084280\relax
\mciteBstWouldAddEndPuncttrue
\mciteSetBstMidEndSepPunct{\mcitedefaultmidpunct}
{\mcitedefaultendpunct}{\mcitedefaultseppunct}\relax
\EndOfBibitem
\bibitem[Zinke-Allmang(1999)]{zinke1999}
Zinke-Allmang,~M. Phase separation on solid surfaces: nucleation, coarsening and coalescence kinetics. \emph{Thin Solid Films} \textbf{1999}, \emph{346}, 1--68\relax
\mciteBstWouldAddEndPuncttrue
\mciteSetBstMidEndSepPunct{\mcitedefaultmidpunct}
{\mcitedefaultendpunct}{\mcitedefaultseppunct}\relax
\EndOfBibitem
\bibitem[Lai \latin{et~al.}(2019)Lai, Han, Spurgeon, Huang, Thiel, Liu, and Evans]{Lai2019}
Lai,~K.~C.; Han,~Y.; Spurgeon,~P.; Huang,~W.; Thiel,~P.~A.; Liu,~D.-J.; Evans,~J.~W. Reshaping, Intermixing, and Coarsening for Metallic Nanocrystals: Nonequilibrium Statistical Mechanical and Coarse-Grained Modeling. \emph{Chemical Reviews} \textbf{2019}, \emph{119}, 6670–6768\relax
\mciteBstWouldAddEndPuncttrue
\mciteSetBstMidEndSepPunct{\mcitedefaultmidpunct}
{\mcitedefaultendpunct}{\mcitedefaultseppunct}\relax
\EndOfBibitem
\bibitem[Mamatkulov and Zhdanov(2025)Mamatkulov, and Zhdanov]{CES2025_OR_Dimers}
Mamatkulov,~M.; Zhdanov,~V.~P. Ostwald ripening of supported metal nanoparticles: Role of dimers and other general trends. \emph{Chemical Engineering Science} \textbf{2025}, \emph{308}, 121373\relax
\mciteBstWouldAddEndPuncttrue
\mciteSetBstMidEndSepPunct{\mcitedefaultmidpunct}
{\mcitedefaultendpunct}{\mcitedefaultseppunct}\relax
\EndOfBibitem
\bibitem[Zhdanov \latin{et~al.}(2012)Zhdanov, Larsson, and Langhammer]{CPL2012_NovelOR_Supported}
Zhdanov,~V.~P.; Larsson,~E.~M.; Langhammer,~C. Novel aspects of Ostwald ripening of supported metal nanoparticles. \emph{Chemical Physics Letters} \textbf{2012}, \emph{533}, 65--69\relax
\mciteBstWouldAddEndPuncttrue
\mciteSetBstMidEndSepPunct{\mcitedefaultmidpunct}
{\mcitedefaultendpunct}{\mcitedefaultseppunct}\relax
\EndOfBibitem
\bibitem[Panciera \latin{et~al.}(2019)Panciera, Tersoff, Gamalski, Reuter, Zakharov, Stach, Hofmann, and Ross]{Panciera2019}
Panciera,~F.; Tersoff,~J.; Gamalski,~A.~D.; Reuter,~M.~C.; Zakharov,~D.; Stach,~E.~A.; Hofmann,~S.; Ross,~F.~M. Surface Crystallization of Liquid Au–Si and Its Impact on Catalysis. \emph{Advanced Materials} \textbf{2019}, \emph{31}, 1806544\relax
\mciteBstWouldAddEndPuncttrue
\mciteSetBstMidEndSepPunct{\mcitedefaultmidpunct}
{\mcitedefaultendpunct}{\mcitedefaultseppunct}\relax
\EndOfBibitem
\bibitem[Stangebye \latin{et~al.}(2022)Stangebye, Lei, Kinghorn, Robertson, Kacher, and Hattar]{AuSi2022_Eutectic_Dynamics}
Stangebye,~S.; Lei,~C.; Kinghorn,~A.; Robertson,~I.; Kacher,~J.; Hattar,~K. Dynamics of the gold–silicon eutectic reaction studied at limited length scales using in situ TEM and STEM. \emph{Journal of Materials Research} \textbf{2022}, \emph{37}, 3842–3854\relax
\mciteBstWouldAddEndPuncttrue
\mciteSetBstMidEndSepPunct{\mcitedefaultmidpunct}
{\mcitedefaultendpunct}{\mcitedefaultseppunct}\relax
\EndOfBibitem
\bibitem[Egerton(2019)]{Egerton2019_Micron_DamageReview}
Egerton,~R. Radiation damage to organic and inorganic specimens in the TEM. \emph{Micron} \textbf{2019}, \emph{119}, 72–87\relax
\mciteBstWouldAddEndPuncttrue
\mciteSetBstMidEndSepPunct{\mcitedefaultmidpunct}
{\mcitedefaultendpunct}{\mcitedefaultseppunct}\relax
\EndOfBibitem
\bibitem[Stangebye \latin{et~al.}(2022)Stangebye, Zhang, Gupta, Zhu, Pierron, and Kacher]{ActaMat2021_EBeamEffects}
Stangebye,~S.; Zhang,~Y.; Gupta,~S.; Zhu,~T.; Pierron,~O.; Kacher,~J. Understanding and quantifying electron beam effects during in situ TEM nanomechanical tensile testing on metal thin films. \emph{Acta Materialia} \textbf{2022}, \emph{222}, 117441\relax
\mciteBstWouldAddEndPuncttrue
\mciteSetBstMidEndSepPunct{\mcitedefaultmidpunct}
{\mcitedefaultendpunct}{\mcitedefaultseppunct}\relax
\EndOfBibitem
\bibitem[Woehl \latin{et~al.}(2020)Woehl, Moser, Evans, and Ross]{MRSBull2020_EBeam_Chemistry}
Woehl,~T.~J.; Moser,~T.; Evans,~J.~E.; Ross,~F.~M. Electron-beam-driven chemical processes during liquid phase transmission electron microscopy. \emph{MRS Bulletin} \textbf{2020}, \emph{45}, 746–753\relax
\mciteBstWouldAddEndPuncttrue
\mciteSetBstMidEndSepPunct{\mcitedefaultmidpunct}
{\mcitedefaultendpunct}{\mcitedefaultseppunct}\relax
\EndOfBibitem
\bibitem[Ramkrishna(2000)]{Ramkrishna2000_PBM}
Ramkrishna,~D. \emph{Population Balances: Theory and Applications to Particulate Systems in Engineering}; Academic Press, 2000\relax
\mciteBstWouldAddEndPuncttrue
\mciteSetBstMidEndSepPunct{\mcitedefaultmidpunct}
{\mcitedefaultendpunct}{\mcitedefaultseppunct}\relax
\EndOfBibitem
\bibitem[Friedlander(2000)]{Friedlander2000_Smoke}
Friedlander,~S.~K. \emph{Smoke, Dust, and Haze: Fundamentals of Aerosol Dynamics}, 2nd ed.; Oxford University Press, 2000\relax
\mciteBstWouldAddEndPuncttrue
\mciteSetBstMidEndSepPunct{\mcitedefaultmidpunct}
{\mcitedefaultendpunct}{\mcitedefaultseppunct}\relax
\EndOfBibitem
\bibitem[Aldous(1999)]{Aldous1999_Coagulation}
Aldous,~D.~J. Deterministic and Stochastic Models for Coalescence (Aggregation and Coagulation): A Review of the Mean-Field Theory for Probabilists. \emph{Bernoulli} \textbf{1999}, \emph{5}, 3--48\relax
\mciteBstWouldAddEndPuncttrue
\mciteSetBstMidEndSepPunct{\mcitedefaultmidpunct}
{\mcitedefaultendpunct}{\mcitedefaultseppunct}\relax
\EndOfBibitem
\bibitem[Marchisio and Fox(2013)Marchisio, and Fox]{MarchisioFox2013_PBM}
Marchisio,~D.~L.; Fox,~R.~O. \emph{Computational Models for Polydisperse Particulate and Multiphase Systems}; Cambridge University Press, 2013\relax
\mciteBstWouldAddEndPuncttrue
\mciteSetBstMidEndSepPunct{\mcitedefaultmidpunct}
{\mcitedefaultendpunct}{\mcitedefaultseppunct}\relax
\EndOfBibitem
\bibitem[Gonzalez and Woods(2006)Gonzalez, and Woods]{gonzales2006}
Gonzalez,~R.~C.; Woods,~R.~E. \emph{Digital Image Processing}, 3rd ed.; Prentice-Hall, Inc.: USA, 2006\relax
\mciteBstWouldAddEndPuncttrue
\mciteSetBstMidEndSepPunct{\mcitedefaultmidpunct}
{\mcitedefaultendpunct}{\mcitedefaultseppunct}\relax
\EndOfBibitem
\bibitem[Holec \latin{et~al.}(2020)Holec, Dumitraschkewitz, Vollath, and Fischer]{Holec2020}
Holec,~D.; Dumitraschkewitz,~P.; Vollath,~D.; Fischer,~F.~D. Surface Energy of Au Nanoparticles Depending on Their Size and Shape. \emph{Nanomaterials} \textbf{2020}, \emph{10}, 484\relax
\mciteBstWouldAddEndPuncttrue
\mciteSetBstMidEndSepPunct{\mcitedefaultmidpunct}
{\mcitedefaultendpunct}{\mcitedefaultseppunct}\relax
\EndOfBibitem
\bibitem[Tyson and Miller(1977)Tyson, and Miller]{tyson1977surface}
Tyson,~W.~R.; Miller,~W.~A. Surface free energies of solid metals: Estimation from liquid surface tension measurements. \emph{Surface Science} \textbf{1977}, \emph{62}, 267--276\relax
\mciteBstWouldAddEndPuncttrue
\mciteSetBstMidEndSepPunct{\mcitedefaultmidpunct}
{\mcitedefaultendpunct}{\mcitedefaultseppunct}\relax
\EndOfBibitem
\bibitem[Yao \latin{et~al.}(1993)Yao, Elder, Guo, and Grant]{Ostwald2D}
Yao,~J.~H.; Elder,~K.~R.; Guo,~H.; Grant,~M. Theory and simulation of Ostwald ripening. \emph{Phys. Rev. B} \textbf{1993}, \emph{47}, 14110--14125\relax
\mciteBstWouldAddEndPuncttrue
\mciteSetBstMidEndSepPunct{\mcitedefaultmidpunct}
{\mcitedefaultendpunct}{\mcitedefaultseppunct}\relax
\EndOfBibitem
\bibitem[Alcock \latin{et~al.}(1984)Alcock, Itkin, and Horrigan]{vapour}
Alcock,~C.~B.; Itkin,~V.~P.; Horrigan,~M.~K. Vapour Pressure Equations for the Metallic Elements: 298–2500K. \emph{Canadian Metallurgical Quarterly} \textbf{1984}, \emph{23}, 309–313\relax
\mciteBstWouldAddEndPuncttrue
\mciteSetBstMidEndSepPunct{\mcitedefaultmidpunct}
{\mcitedefaultendpunct}{\mcitedefaultseppunct}\relax
\EndOfBibitem
\bibitem[Chandler(1987)]{chandler1987}
Chandler,~D. \emph{Introduction to Modern Statistical Mechanics}; Oxford University Press: New York, 1987\relax
\mciteBstWouldAddEndPuncttrue
\mciteSetBstMidEndSepPunct{\mcitedefaultmidpunct}
{\mcitedefaultendpunct}{\mcitedefaultseppunct}\relax
\EndOfBibitem
\end{mcitethebibliography}

\clearpage
\onecolumn

\renewcommand{\theequation}{S\arabic{equation}}
\setcounter{equation}{0}
\renewcommand{\thefigure}{S\arabic{figure}}
\setcounter{figure}{0}
\setcounter{page}{1}
\section{Supporting Information}

\subsection{Diffusion-mediated sublimation of supported nanoparticles}

\begin{figure}[h!]
\centering   
\includegraphics[width=0.8\linewidth]{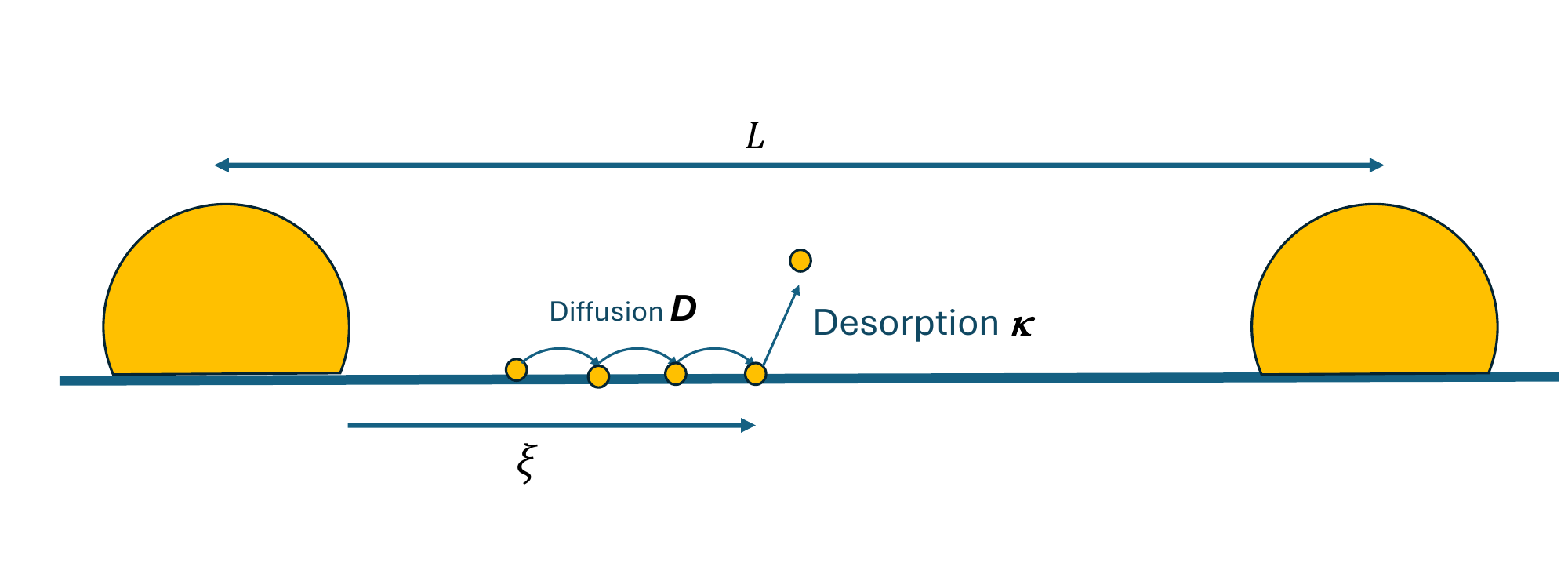}
\caption{\textbf{Schematic of substrate-mediated evaporation.}
  Supported NP of radius $R$ on a substrate with adatom diffusion length $\xi$ and interparticle spacing $L$. When $L\gtrsim \xi$, particles behave quasi-independently; evaporation includes (i) direct sublimation from the NP surface and (ii) desorption of adatoms generated within a substrate ring of order $\xi$ around each NP. The latter yields an approximately size-independent contribution to the per-particle mass loss.
}\label{fig:model1}
\end{figure}

Here we develop a two-scale, self-consistent theory for sublimation of nanoparticles (NPs) supported on a substrate. Atoms detach from an NP, diffuse as adatoms on the substrate, and are removed by desorption or recaptured by neighboring particles. Two channels contribute to mass loss: (i) direct evaporation from the exposed NP surface and (ii) substrate-mediated exchange via the adatom field (as schematically shownm in  Fig. \ref{fig:model1}).

Throughout, \(C(\mathbf x,t)\) denotes the \emph{adatom number density per unit area} on the substrate \([\#/\mathrm{area}]\). The 2D diffusion coefficient is \(D_a\) and the desorption rate is \(Q\), so that in the absence of particles
\begin{equation}
\partial_t C = D_a\nabla^2 C - Q C.
\label{eq:bare}
\end{equation}
The associated bare screening length is
\begin{equation}
\xi_0=\sqrt{\frac{D_a}{Q}}.
\label{eq:xi0}
\end{equation}

Particle \(i\) has radius \(R_i\) and contact angle \(\theta\); its contact-line radius is \(R_i\sin\theta\). At the NP edge, we assume the 2D adatom concentration \(C_{\mathrm{eq}}(R_i)\) to be in equilibrium with the NP. This concentration must correspond to the same mean evaporation rate per unit area \(U\) as from the NP surface:
\begin{equation}
C\Big |_{|\mathbf x-\mathbf x_i|=R_i \sin \theta}
= C_{\mathrm{eq}}(R_i)
= \frac{U}{Q v_a}\exp\!\left(\frac{R_\gamma}{R_i}\right).
\end{equation}
Here \(R_\gamma=2\gamma v_a/(k_BT)\) is the capillary length (\(\gamma\) is the surface energy and \(v_a\) the atomic volume). 
Note that \(U\) has units of \([\mathrm{volume}/(\mathrm{area}\cdot \mathrm{time})] = \mathrm{nm/s}\), i.e., it has the physical meaning of an effective evaporation velocity from a flat interface.

\subsubsection{Coarse-grained description}

On coarse scales, each NP is represented as a localized drain/source center at \(\mathbf x_i\). The coarse-grained equation for \(C(\mathbf x,t)\) is
\begin{equation}
\partial_t C(\mathbf x,t)
=
D_a\nabla^2 C - Q C
-
\sum_i 2\pi \kappa_i D_a \bigl[C_{\mathrm{eq}}(R_i)-C(\mathbf x_i)\bigr] \delta(\mathbf x-\mathbf x_i),
\label{eq:cg_delta}
\end{equation}
where \(\kappa_i\) is a dimensionless geometric factor (determined from the near-field solution below). With this convention, the \emph{total} atom flux from particle \(i\) to the substrate is
\[
\dot N_{{\rm sub},i}=2\pi D_a\kappa_i\bigl[C_{\mathrm{eq}}(R_i)-C(\mathbf x_i)\bigr].
\]

Coarse-graining over an ensemble of areal density \(n\) yields
\begin{equation}
\partial_t C
=
D_a\nabla^2 C
-
\Bigl(Q + 2\pi D_a n \langle \kappa\rangle\Bigr) C
+
2\pi D_a n \langle \kappa\, C_{\mathrm{eq}}(R)\rangle,
\label{eq:cg_avg}
\end{equation}
where \(\langle\cdot\rangle\) denotes averaging over the size distribution.

\paragraph{Renormalized screening length.}
Coarse-grained fluctuations relax with effective rate
\[
Q_{\mathrm{eff}} \equiv Q + 2\pi D_a n \langle\kappa\rangle,
\]
so the ensemble-renormalized screening length is
\begin{equation}
\boxed{ \;
\xi \equiv \sqrt{\frac{D_a}{Q_{\mathrm{eff}}}}
\quad\Longrightarrow\quad
\frac{1}{\xi^2}=\frac{1}{\xi_0^2}+2\pi n\langle\kappa\rangle.
\; }
\label{eq:xi_def}
\end{equation}

\paragraph{Uniform mean-field background.}
For a spatially uniform steady state \(C(\mathbf x)\equiv C_*\), Eq.~\eqref{eq:cg_avg} gives
\begin{equation}
\boxed{ \;
C_* = \frac{2\pi D_a n \langle \kappa\, C_{\mathrm{eq}}(R)\rangle}{Q + 2\pi D_a n\langle\kappa\rangle}
= \frac{2\pi \xi^2 U n \langle \kappa\,\chi\rangle }{v_a Q}.
\; }
\label{eq:Cstar}
\end{equation}

\subsubsection{Near-field concentration profile about a tagged particle}

After obtaining \(C_*\) and \(\xi\) from the coarse-grained analysis, we consider the near-field around a tagged particle \(i\). Let
\[
r\equiv|\mathbf x-\mathbf x_i|,\qquad C(r)=C_*+c(r),
\]
where \(C(r)\) is the overall concentration profile and \(c(r)\) is the localized perturbation.

In the quasi-static approximation, \(C(r)\) satisfies the screened diffusion equation with \(Q_{\rm eff}=D_a/\xi^2\):
\begin{equation}
D_a\left(\frac{1}{r}\frac{\partial}{\partial r}r\frac{\partial}{\partial r}\right)C(r) - Q_{\mathrm{eff}}\,C(r)=0,
\label{eq:near_full}
\end{equation}
Equivalently, the perturbation \(c(r)=C(r)-C_*\) satisfies
\begin{equation}
\left(\frac{1}{r}\frac{\partial}{\partial r}r\frac{\partial}{\partial r}\right)c(r) - \frac{1}{\xi^2}\,c(r)=0,
\label{eq:near_pert}
\end{equation}
with boundary conditions
\[
c(r=R_i\sin\theta)=C_{\mathrm{eq}}(R_i)-C_*,
\qquad
c(r\to\infty)=0.
\]
The unique radially decaying solution is
\begin{equation}
c(r)=\bigl[C_{\mathrm{eq}}(R_i)-C_*\bigr]\frac{K_0(r/\xi)}{K_0(x_i)},
\qquad x_i\equiv\frac{R_i\sin\theta}{\xi}.
\label{eq:c_solution}
\end{equation}
Thus the overall near-field profile is
\begin{equation}
\boxed{\;
C(r)=C_*+\bigl[C_{\mathrm{eq}}(R_i)-C_*\bigr]\frac{K_0(r/\xi)}{K_0(x_i)}.
\;}
\label{eq:C_profile}
\end{equation}

The radial flux density at the contact line is
\[
j_i \equiv -D_a\left.\frac{\partial C}{\partial r}\right|_{r=R_i\sin\theta}
=
\frac{D_a}{\xi}\bigl[C_{\mathrm{eq}}(R_i)-C_*\bigr]\frac{K_1(x_i)}{K_0(x_i)}.
\]
Multiplying by the perimeter \(2\pi R_i \sin\theta\) gives the total substrate-mediated atom flux from the NP:
\begin{equation}
\boxed{ \;
\dot N_{{\rm sub},i}
= 2\pi D_a\,\kappa_i\bigl[C_{\mathrm{eq}}(R_i)-C_*\bigr],
\qquad
\kappa_i \equiv x_i\frac{K_1(x_i)}{K_0(x_i)}.
\; }
\label{eq:Nsub_kappa}
\end{equation}

For \(x_i\ll 1\) (i.e., \(R_i\sin\theta\ll\xi\)), the kernel has the two-dimensional logarithmic form
\begin{equation}
\kappa_i \approx \frac{1}{\ln\!\bigl(\xi/(R_i\sin\theta)\bigr)}.
\label{eq:kappa_small}
\end{equation}

\subsubsection{Total shrinkage law with Ostwald correction}

Direct evaporation from the exposed particle surface provides a volumetric loss proportional to surface area:
\[
\dot N_{{\rm surf},i} = U\,2\pi R_i^2(1-\cos\theta)\,\chi_i,
\qquad \chi_i\equiv\chi(R_i).
\]
Combining substrate-mediated exchange (Eq.~\eqref{eq:Nsub_kappa}) with surface evaporation and rewriting in terms of \(\xi\) gives the compact volumetric shrinkage law:
\begin{equation}
\boxed{ \;
\dot V_i
=
-2\pi U \xi^2
\Bigg[
\chi_i\left(1-\cos\theta\right)\left(\frac{R_i}{\xi}\right)^2
+
\kappa_i\left(
\chi_i
+
2\pi n \xi_0^2\left(\langle\kappa\rangle\chi_i-\langle\kappa\chi\rangle\right)
\right)
\Bigg].
\; }
\label{eq:Vdot_box}
\end{equation}

\begin{figure}[h!]
\centering   
\includegraphics[width=0.8\linewidth]{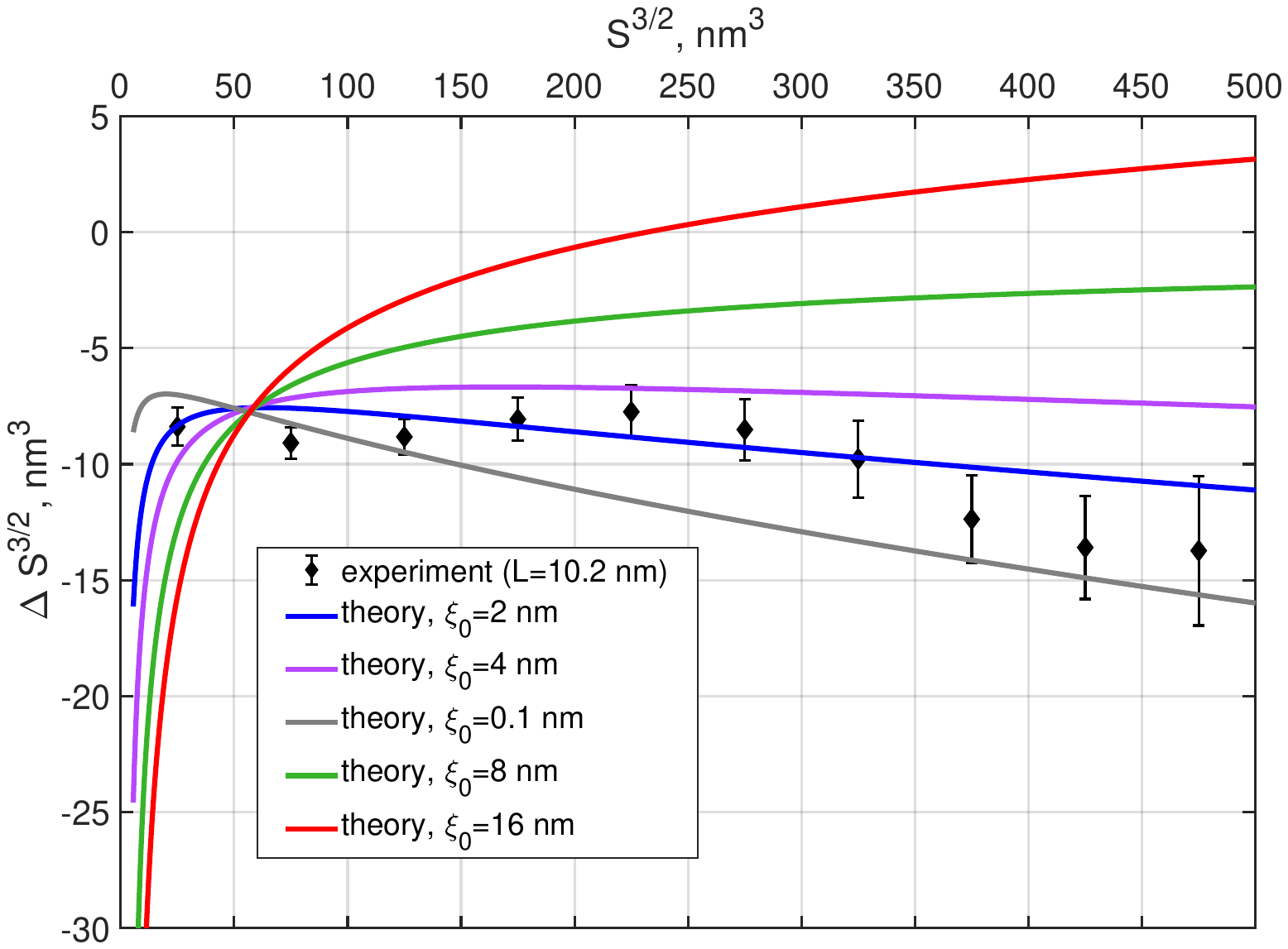}
\caption{Experimentally observed mean sublimation rate $\dot V$, and  model predictions for various values of bare screening length $\xi_0$
}\label{fig:scatter}
\end{figure}

Equation~\eqref{eq:Vdot_box} separates the shrinkage rate into three physically distinct contributions.

\emph{(i) Direct surface evaporation.}
The first term scales with surface area (\(\propto R_i^2\)) and is amplified by the curvature factor \(\chi_i\).

\emph{(ii) Substrate-mediated evaporation.}
The term \(\kappa_i\chi_i\) represents volume loss through substrate diffusion and desorption in the isolated-particle limit.

\emph{(iii) Collective (Ostwald) correction.}
The final term accounts for the self-consistent background generated by the ensemble. Its magnitude is controlled by the dimensionless density parameter \(2\pi n\xi_0^2\). The statistical factor \(\langle\kappa\rangle\chi_i-\langle\kappa\chi\rangle\) vanishes for a monodisperse ensemble and is nonzero only for polydispersity.

Together, these terms describe a crossover from isolated surface-controlled sublimation to diffusion-limited, collectively screened kinetics as particle density and polydispersity increase. Theoretical predictions for various values of $\xi_0$ are presented in Figure \ref{fig:scatter}, alongside with the experimentally measured sublimation rate.

\subsection{Shot noise, shape fluctuations, and proxy statistics}

\subsubsection{Langevin description of volume fluctuations}

Experimentally measured size proxies (e.g., \(S^{3/2}\), where \(S\) is projected area) differ from the true particle volume because of shape fluctuations. We write
\[
V_i(t)=f_i(t)\,S_i^{3/2}(t),
\]
where \(f_i(t)\) is a slowly varying geometric factor. On times longer than the shape-relaxation time, \(f_i\) may be treated as approximately constant, and the stochastic evolution of the particle volume is described by
\begin{equation}
\dot V_i(t)=\langle\dot V\rangle_{R_i}+\eta_i(t),
\label{eq:langevin}
\end{equation}
where \(\eta_i(t)\) is a zero-mean gaussian noise term. that  satisfies
\[
\langle \eta_i(t)\eta_i(t')\rangle
=
\Lambda\,\delta(t-t').
\]

For time increments \(\Delta t\) longer than the noise correlation time,
\begin{equation}
\mathrm{var}\!\left[V(t+\Delta t)-V(t)\right]
=
\Lambda\,\Delta t.
\label{eq:var_linear}
\end{equation}

\subsubsection{Shot-noise estimate of the noise amplitude}

We estimate \(\Lambda\) by treating atomic attachment/detachment as a Poisson process. Each atomic event changes the particle volume by an amount \(v_a\) (atomic volume), and events occur at rate \(\Gamma\) [1/time]. For Poisson statistics,
\begin{equation}
\boxed{\;
\Lambda \sim v_a^2\,\Gamma,
\;}
\label{eq:Lambda_poisson}
\end{equation}

To estimate \(\Gamma\), we use the substrate-mediated flux near the particle edge. The characteristic adatom concentration scale is
\[
C_{\rm eq}(R) \sim C_{\mathrm{eq}}(R)
=
\frac{U}{Q v_a}\,e^{R_\gamma/R}.
\]
The diffusive particle–substrate exchange scvale as
\begin{equation}
\Gamma
\sim
2\pi R\,D_a\,\frac{C_{\rm eq}(R)}{a}
\label{eq:Gamma_est}
\end{equation}
Here $a\sim v_a^{1/3}$ is atomic scale. 

Substituting \(C_{\rm eq}(R)\) gives
\[
\Gamma
\sim
\frac{2\pi D_a U}{Q v_a^{4/3}}\,
\langle R\,e^{R_\gamma/R}\rangle.
\]

Using \(D_a/Q=\xi_0^2\), and Eq.~\eqref{eq:Lambda_poisson} yields

\begin{equation}
\boxed{\;
\Lambda
\sim
2\pi \xi_0^2 U\,v_a^{2/3}\,
\langle R\,e^{R_\gamma/R}\rangle.
\;}
\label{eq:Lambda_final}
\end{equation}

Both the deterministic shrinkage rate (Eq.~\eqref{eq:Vdot_box}) and the fluctuation amplitude \(\Lambda\) depend on the same kinetic parameters \(D_a\), \(Q\), \(U\), and on the curvature factor \(e^{R_\gamma/R}\). Thus the mean drift and the variance growth originate from the same underlying diffusion–desorption physics.

\end{document}